%% file: ms.tex
\documentclass[iop]{emulateapj}

\usepackage{natbib}
\usepackage{amsmath}
\usepackage{epsfig}
\usepackage{graphicx}
\usepackage{natbib}
\usepackage{xspace}
\usepackage{longtable}
\usepackage[T1]{fontenc}
\usepackage{topcapt}
\usepackage[super]{nth}
\newcommand{\hlf}{\tfrac{1}{2}}
\newcommand{\tdur}{\ensuremath{\Delta T}\xspace}
\newcommand{\ep}{\ensuremath{t_{0}}\xspace}
\newcommand{\Per}{\ensuremath{P}\xspace}

\newcommand{\df}{\delta F}
\newcommand{\mdf}{\overline{\df}}

\newcommand{\teff}{\ensuremath{ {\rm{T}_{\rm eff}} }\xspace}
\newcommand{\logg}{\ensuremath{ {\log g }}\xspace}

\newcommand{\SNRC}{\ensuremath{\text{SNR}_\text{CDPP}}\xspace}
\newcommand{\NSNRnoteKOI}{609\xspace}
\newcommand{\NSNR}{738\xspace}
 \newcommand{\NSNRnotP}{347\xspace} \newcommand{\NnotSolarSubset}{10\xspace} 
\newcommand{\Np}{\ensuremath{n_{\text{pl,cell}}}\xspace}
\newcommand{\NpAug}{\ensuremath{n_{\text{pl,aug,cell}}}}
\newcommand{\fcell}{\ensuremath{f_{\text{cell}}}\xspace }
\newcommand{\fcellBa}{\ensuremath{f_{\text{cell,Batalha}}}\xspace}
\newcommand{\flogA}{\ensuremath{d^{2}\fcell / d\log{P} / d\log{R_P}}\xspace }
\newcommand{\NstarAmen}{ n_{\star,\text{amen}} }

\newcommand{\Rsun}{\ensuremath{R_{\odot}}\xspace }
\newcommand{\fonetwo}{\ensuremath{15.1^{+1.8}_{-2.7}\%}\xspace}
\newcommand{\Kepler}{\textit{Kepler}\xspace} 
 
\newcommand{\SpecMatch}{{\tt SpecMatch}\xspace}
\newcommand{\TERRA}{{\tt TERRA}\xspace}

\renewcommand{\Re}{\ensuremath{ R_{E} }\xspace} 
\newcommand{\Rp}{\ensuremath{ R_{P} }\xspace}
\newcommand{\Rstar}{\ensuremath{R_{\star}}\xspace} 
\newcommand{\rrat}{\ensuremath{\Rp / \Rstar}\xspace}  
\newcommand{\rratfrac}{\ensuremath{\frac{\Rp}{\Rstar}}\xspace}  

\begin{document}
\shortauthors{E. Petigura \& G. Marcy}

\title{A Plateau in the Planet Population Below Twice the Size of Earth}

\author{Erik~A.~Petigura}
\affil{Astronomy Department, University of California, Berkeley, CA 94720, USA}
\email{epetigura@berkeley.edu}
\author{Geoffrey~W.~Marcy}
\affil{Astronomy Department, University of California, Berkeley, CA 94720, USA}
\author{Andrew~W.~Howard}
\affil{Institute for Astronomy, University of Hawaii, 2680 Woodlawn Drive, Honolulu, HI 96822, USA}


\begin{abstract}
We carry out an independent search of \Kepler photometry for small transiting planets with sizes 0.5--8.0 times that of Earth and orbital periods between 5 and 50 days, with the goal of measuring the fraction of stars harboring such planets. We use a new transit search algorithm, \TERRA, optimized to detect small planets around photometrically quiet stars. We restrict our stellar sample to include the 12,000 stars having the lowest photometric noise in the \Kepler survey, thereby maximizing the detectability of Earth-size planets. We report 129 planet candidates having radii less than 6 \Re found in 3 years of \Kepler photometry (quarters 1--12). Forty-seven of these candidates are not in \cite{Batalha12}, which only analyzed photometry from quarters 1--6. We gather Keck HIRES spectra for the majority of these targets leading to precise stellar radii and hence precise planet radii. We make a detailed measurement of the completeness of our planet search. We inject synthetic dimmings from mock transiting planets into the actual \Kepler photometry. We then analyze that injected photometry with our \TERRA pipeline to assess our detection completeness for planets of different sizes and orbital periods. We compute the occurrence of planets as a function of planet radius and period, correcting for the detection completeness as well as the geometric probability of transit, $\Rstar/a$. The resulting distribution of planet sizes exhibits a power law rise in occurrence from 5.7 \Re down to 2 \Re, as found in \cite{Howard12}. That rise clearly ends at 2 \Re. The occurrence of planets is consistent with constant from 2 \Re toward 1 \Re. This unexpected plateau in planet occurrence at 2 \Re suggests distinct planet formation processes for planets above and below 2 \Re. We find that $15.1^{+1.8}_{-2.7}$\% of solar type stars---roughly one in six---has a 1--2~\Re planet with \Per~=~5--50~days.
\end{abstract}

\section{Introduction}

The \Kepler Mission has discovered an extraordinary sample of more than 2300 planets with radii ranging from larger than Jupiter to smaller than Earth \citep{Borucki11,Batalha12}. Cleanly measuring and debiasing this distribution will be one of {\em Kepler's} great legacies. \cite{Howard12}, H12 hereafter, took a key step, showing that the planet radius distribution increases substantially with decreasing planet size down to at least 2 \Re. While the distribution of planets of all periods and radii contains a wealth of information, we choose to focus on the smallest planets.  Currently, only \Kepler is able to make quantitative statements about the occurrence of planets down to 1 \Re.

The occurrence distributions in H12 were based on planet candidates detected in the first four months of \Kepler photometry \citep{Borucki11}. These planet candidates were detected by a sophisticated pipeline developed by the \Kepler team Science Operations Center \citep{Twicken10,Jenkins10}.\footnote{Since H12, \cite{Batalha12} added many candidates, bringing the number of public KOIs to >~2300. In addition, the \Kepler team planet search pipeline has continued to evolve \citep{Smith12,Stumpe12}.}
Understanding pipeline completeness, the fraction of planets missed by the pipeline as a function of size and period, is a key component to measuring planet occurrence. Pipeline completeness can be assessed by injecting mock dimmings into photometry and measuring the rate at which injected signals are found. The completeness of the official \Kepler pipeline has yet to be measured in this manner. This was the key reason why H12 were cautious interpreting planet occurrence under 2 \Re.

In this work, we focus on determining the occurrence of small planets. To maximize our sensitivity to small planets, we restrict our stellar sample to include only the 12,000 stars having the lowest photometric noise in the \Kepler survey. We comb through quarters 1--12 (Q1--Q12) --- 3 years of \Kepler photometry --- with a new algorithm, \TERRA, optimized to detect low signal-to-noise transit events. We determine \TERRA's sensitivity to planets of different periods and radii by injecting synthetic transits into \Kepler photometry and measuring the recovery rate as a function of planet period and radius. 

We describe our selection of 12,000 low-noise targets in Section~\ref{sec:Sample}. We comb their photometry for exoplanet transits with \TERRA, introduced in Section~\ref{sec:TERRA}. We report candidates found with \TERRA (Section~\ref{sec:TERRAplanetYield}), which we combine with our measurement of pipeline completeness (Section~\ref{sec:MC}) to produce debiased measurements of planet occurrence (Section~\ref{sec:occurrenceme}). We offer some comparisons between \TERRA planet candidates and those from \cite{Batalha12} in Section~\ref{sec:compare} as well as occurrence measured using both catalogs in Section~\ref{sec:OccurCOMB}. We offer some interpretations of the constant occurrence rate for planets smaller than 2 \Re in Section~\ref{sec:Discussion}.

\section{The Best12k Stellar Sample}
\label{sec:Sample}
We restrict our study to the best 12,000 solar type stars from the perspective of detecting transits by Earth-size planets, hereafter, the ``Best12k'' sample. For the smallest planets, uncertainty in the occurrence distribution stems largely from pipeline incompleteness due to the low signal-to-noise ratio (SNR) of an Earth-size transit. 

\begin{figure}
\includegraphics[width=0.48\textwidth]{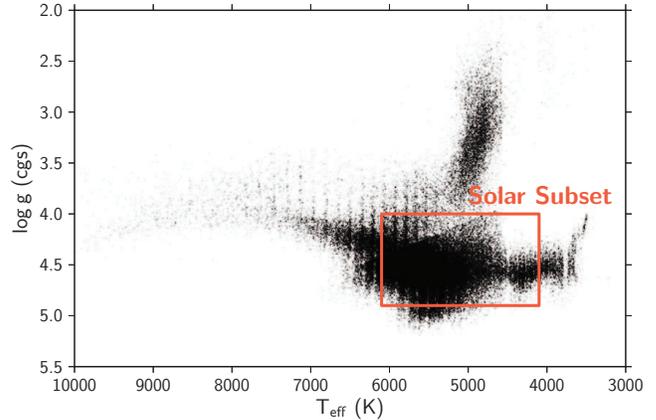}
\caption{\Kepler target stars observed every quarter from Q1--Q9.  The rectangle marks the ``solar subset'' of stars with \teff~=~4100--6100~K and \logg~=~4.0--4.9 (cgs).}
\label{fig:solarsubset}
\end{figure}

Our initial sample begins with the 102,835 stars that were observed during every quarter from Q1--Q9.\footnote{We ran \TERRA on Q1-Q12 photometry, but we selected the Best12k sample before Q10-Q12 were available.} 
From this sample, following H12, we select 73,757 ``solar subset'' stars that are solar-type G and K having \teff = 4100--6100~K and \logg = 4.0--4.9~(cgs). \teff and \logg values are present in the \Kepler Input Catalog (KIC; \citealp{Brown11}) which is available online.\footnote{http://archive.stsci.edu/Kepler/kic.html}
Figure~\ref{fig:solarsubset} shows the KIC-based \teff and \logg values as well as the solar subset. KIC stellar parameters have large uncertainties: $\sigma(\logg) \sim$ 0.4~dex and $\sigma(\teff) \sim$ 200~K \citep{Brown11}. As we will discuss in Section~\ref{sec:TERRAplanetYield}, we determine stellar parameters for the majority of \TERRA planet candidates spectroscopically. For the remaining cases, we use stellar parameters that were determined photometrically, but incorporated a main sequence prior \citep{Batalha12}. After refining the stellar parameters, we find that \NnotSolarSubset of the 129 \TERRA planet candidates fall outside of the \teff = 4100--6100~K and \logg = 4.0--4.9~(cgs) solar subset.

\begin{figure}[htbp]
\includegraphics[width=\columnwidth]{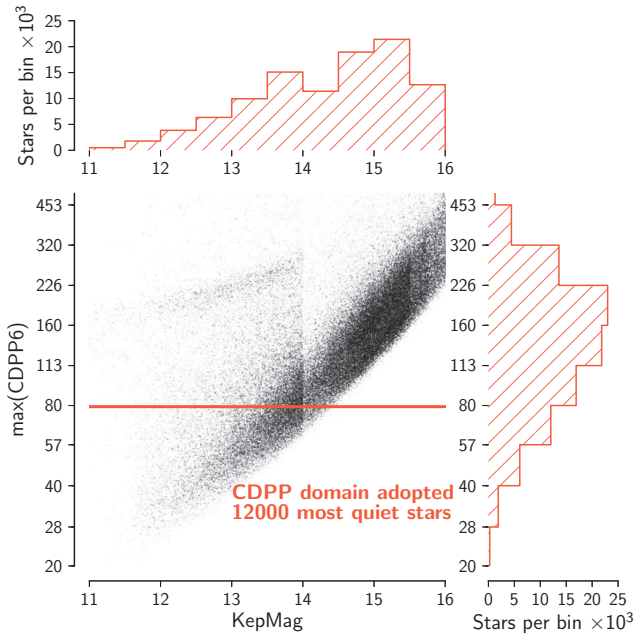}
\caption{Stellar photometric noise level plotted against \Kepler magnitude. Noise level is the maximum value of CDPP6 over Q1--Q9. Of the 73,757 stars that pass our cuts on \teff and \logg, we select the 12,000 most quiet stars. The line shows max(CDPP6) = 79.2~ppm, corresponding to the noisiest star in the Best12k sample, well below the median value of 143~ppm.}
\label{fig:noise}
\end{figure}

From the 73,757 stars that pass our cuts on \logg and \teff, we choose the 12,000 lowest noise stars. \Kepler target stars have a wide range of noise properties, and there are several ways of quantifying photometric noise. The \Kepler team computes quantities called CDPP3, CDPP6, and CDPP12, which are measures of the photometric scatter in 3, 6, and 12 hour bins \citep{Jenkins10}. Since CDPP varies by quarter, we adopt the maximum 6-hour CDPP over Q1--Q9 as our nominal noise metric. We use the maximum noise level (as opposed to median or mean) because a single quarter of noisy photometry can set a high noise floor for planet detection. One may circumvent this problem by removing noisy regions of photometry, which is a planned upgrade to \TERRA. Figure~\ref{fig:noise} shows the distribution of max(CDPP6) among the 73,757 stars considered for our sample. 

In choosing our sample, we wanted to include stars amenable to the detection of planets as small as 1 \Re. We picked the 12,000 quietest stars based on preliminary completeness estimates. The noisiest star in the Best12k sample has max(CDPP6) of 79.2~ppm. We estimated that the $\sim$ 100~ppm transit of an Earth-size planet would be detected at $\SNRC \sim 1.25$.\footnote{\SNRC, the expected SNR using the max(CDPP6) metric, is different from the SNR introduced in Section~\ref{sec:TERRAgrid}. \SNRC is more similar to the SNR computed by the \Kepler team, which adopts \SNRC > 7.1 as their detection threshold.}
Given that Q1-Q12 contains roughly 1000 days of photometry, we expected to detect a 5-day planet at $ \SNRC  \sim 1.25 \times \sqrt{1000/5} \sim 18$ (a strong detection) and to detect a 50-day planet at $\SNRC  \sim 1.25 \times \sqrt{1000/50} \sim 5.6$ (a marginal detection). In our detailed study of completeness, described in Section~\ref{sec:MC}, we find that \TERRA recovers most planets down to 1 \Re having \Per~=~5--50~days.

We draw stars from the H12 solar subset for two reasons. First, we may compare our planet occurrence to that of H12 without the complication of varying occurrence with different stellar types. We recognize that subtle differences may exist between the H12 and Best12k stellar sample. One such difference is that the Best12k is noise-limited, while the H12 sample is magnitude-limited. H12 included bright stars with high photometric variability, which are presumably young and/or active stars. Planet formation efficiency could depend on stellar age. Planets may be less common around older stars that formed before the metallicity of the Galaxy was enriched to current levels. This work assesses planet occurrence for a set of stars that are systematically selected to be 3-10 Gyr old by virtue of their reduced magnetic activity.

The second reason for adopting the H12 solar subset is a practical consideration of our completeness study. As shown in Section~\ref{sec:MC}, we parameterize pipeline efficiency as a function of \Per and \Rp. Because M-dwarfs have smaller radii than G-dwarfs, an Earth-size planet dims an M-dwarf more substantially and should be easier for \TERRA to detect. Thus, measuring completeness as a function of \Per, \Rp, {\em and} \Rstar (or perhaps \Per and \Rp/\Rstar) is appropriate when analyzing stars of significantly different sizes. Such extensions are beyond the scope of this paper, and we consider stars with $\Rstar \sim \Rsun$.

\section{Planet Search Pipeline}
\label{sec:TERRA}
Identifying the smallest transiting planets in \Kepler photometry requires a sophisticated automated pipeline. Our pipeline is called ``\TERRA'' and consists of three major components. First, \TERRA calibrates photometry in the time domain. Then, \TERRA combs the calibrated photometry for periodic, box-shaped signals by evaluating the signal-to-noise ratio (SNR) over a finely-spaced grid in transit period (\Per), epoch (\ep) and duration (\tdur). Finally, \TERRA fits promising signals with a \cite{Mandel02} transit model and rejects signals that are not consistent with an exoplanet transit. We review the calibration component in Section~\ref{sec:TERRAcal}, but refer the reader to \cite{Petigura12} for a detailed description. We present, for the first time, the grid-search and light curve fitting components in Sections~\ref{sec:TERRAgrid} and \ref{sec:TERRAdv}.

\subsection{Photometric Calibration}
\label{sec:TERRAcal}
We briefly review the major time domain components of \TERRA; for a more complete description, please refer to \cite{Petigura12}.  We begin with \Kepler ``simple aperture long cadence photometry,'' which we downloaded from the Mikulski Archive for Space Telescopes (MAST). This photometry is the total photoelectrons accumulated within a predefined target aperture over a 29.4 minute interval \citep{KeplerArchiveManual}. We remove thermal settling events manually and cosmic rays using a median filter.  Next, we remove photometric trends longer than 10 days with a high-pass filter.  Finally, we identify photometric modes shared by a large ensemble of stars with using a robust principal components analysis. The optimum linear combination of the four most significant modes is removed from each light curve individually.

\subsection{Grid-based Transit Search}
\label{sec:TERRAgrid}

We then search for periodic, box-shaped signals in ensemble-calibrated photometry. Such a search involves evaluating the SNR over a finely sampled grid in period (\Per), epoch (\ep), and duration (\tdur), i.e.
 \begin{equation}
\text{SNR} = \text{SNR}(\Per,\ep,\tdur).
\end{equation}
Our approach is similar to the widely-used {\tt BLS} algorithm of \cite{Kovacs02} as well as to the {\tt TPS} component of the \Kepler pipeline \citep{Jenkins10TPS}. {\tt BLS}, {\tt TPS}, and \TERRA are all variants of a ``matched filter'' (North 1943). The way in which such an algorithm searches through \Per, \ep, and \tdur is up to the programmer. We choose to search first through \tdur (outer loop), then \Per, and finally, \ep (inner loop).

For computational simplicity, we consider transit durations that are integer numbers of long cadence measurements. Since we search for transits with \Per~=~5--50 days, we try $\tdur = [3,5,7,10,14,18]$ long cadence measurements, which span the range of expected transit durations, 1.5 to 8.8 hours, for G and K dwarf stars.

After choosing \tdur, we compute the mean depth, $\mdf(t_{i})$, of a putative transit with duration = \tdur centered at $t_{i}$ for each cadence. $\mdf$ is computed via

\begin{equation}
\mdf(t_{i}) = \sum_{j} F(t_{i-j}) G_{j}
\end{equation}
where $F(t_{i})$ is the median-normalized stellar flux at time $t_{i}$ and $G_{j}$ is the $j$th element of the following kernel
\begin{equation}
\mathbf{G} = \frac{1}{\tdur}
     \left[
          \underbrace{\hlf, \dots, \hlf}_{\tdur},
          \underbrace{-1, \dots, -1}_{\tdur},
          \underbrace{\hlf, \dots, \hlf}_{\tdur}
     \right].
\end{equation}
As an example, if $\tdur = 3$, 

\begin{equation}
\mathbf{G} = \frac{1}{3}
     \left[
          \hlf, \hlf,\hlf ,-1,-1, -1,\hlf, \hlf,\hlf
     \right].
\end{equation}

We search over a finely sampled grid of trial periods from 5--50 days and epochs ranging from $t_{\text{start}}$ to $t_{\text{start}}+\Per$, where $t_{\text{start}}$ is the time of the first photometric observation. For a given (\Per,\ep,\tdur) there are $N_T$ putative transits with depths $\mdf_{i}$, for $i = 0, 1, \hdots, N_T-1$. For each (\Per,\ep,\tdur) triple, we compute SNR from
\begin{equation}
\text{SNR} = \frac{\sqrt{N_T}}{\sigma} \text{mean}(\mdf_i),
\label{eqn:SNR}
\end{equation}
where $\sigma$ is a robust estimate (median absolute deviation) of the noise in bins of length \tdur.

\newcommand{\PcadO}{\ensuremath{ P_{\text{cad,0}}}\xspace}

For computational efficiency, we employ the ``Fast Folding Algorithm'' (FFA) of \cite{Staelin69} as implemented in Petigura \& Marcy (2013; in prep.). Let \PcadO be a trial period that is an integer number of long cadence measurements, e.g. \PcadO = 1000 implies $\Per = 1000\times 29.4~\text{min}$ = 20.43~days. Let $N_{\text{cad}} = 51413$ be the length of the Q1-Q12 time series measured in long cadences. Leveraging the FFA, we compute SNR at the following periods:
\begin{equation}
P_{\text{cad},i} = \PcadO + \frac{i}{M-1}; \; i=0, 1, \hdots, M-1
\end{equation}
where $M = N_{\text{cad}} / \PcadO $ rounded up to the nearest power of two. In our search from 5--50~days, \PcadO ranges from 245--2445, and we evaluate SNR at $\sim10^5$ different periods. At each $P_{\text{cad,i}}$ we evaluate SNR for \PcadO different starting epochs. All told, for each star, we evaluate SNR at $\sim 10^9$ different combinations of \Per, \ep, and \tdur.

\begin{figure*}
\includegraphics[width=\textwidth]{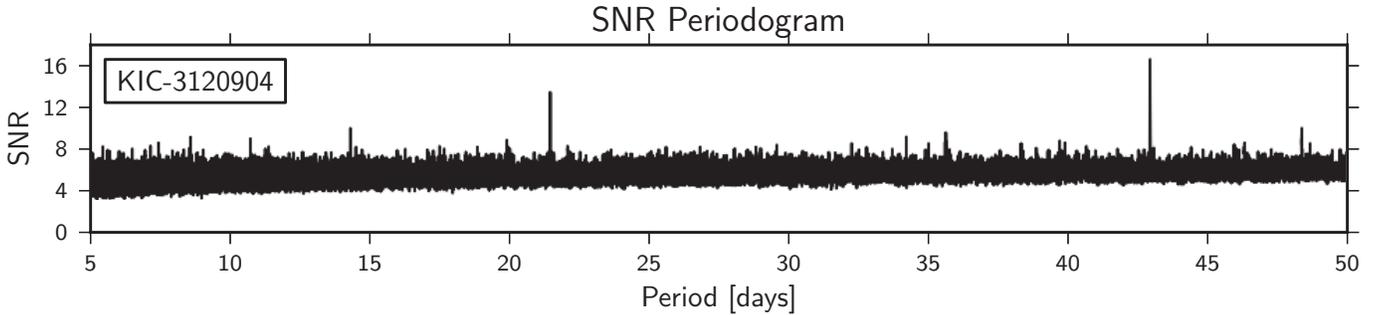}
\caption{SNR periodogram of KIC-3120904 photometry. We evaluate SNR over a finely-spaced, three-dimensional grid of \Per, \ep, and \tdur. We store the maximum SNR for each trial period, resulting in a one-dimensional distribution of SNR. A planet candidate (not in \citealt{Batalha12}) produces a SNR peak of 16.6 at \Per~=~42.9~days, which rises clearly above the detection floor of SNR $\sim$ 8.}
\label{fig:SNR-Periodogram}
\end{figure*}

Due to runtime and memory constraints, we store only one SNR value for each of the trial periods. \TERRA stores the maximum SNR at that period for all \tdur and \ep. We refer to this one-dimensional distribution of SNR as the ``SNR periodogram,'' and we show the KIC-3120904 SNR periodogram in Figure~\ref{fig:SNR-Periodogram} as an example. Because we search over many \tdur and \ep at each trial period, fluctuations often give rise SNR $\sim$ 8 events and set the detectability floor in the SNR periodogram. For KIC-3120904, a star not listed in the \cite{Batalha12} planet catalog, we see a SNR peak of 16.6, which rises clearly above stochastic background.

If the maximum SNR in the SNR periodogram exceeds 12, we pass that particular (\Per,\ep,\tdur) on to the ``data validation'' (DV) step, described in the following section, for additional vetting. We chose 12 as our SNR threshold by trial and error. Note that the median absolute deviation of many samples drawn from a Gaussian distribution is 0.67 times the standard deviation, i.e. $\sigma_{\rm{MAD}} = 0.67 \sigma_{\rm{STD}}$. Therefore, \TERRA SNR = 12 corresponds roughly to SNR = 8 in a {\tt BLS} or {\tt TPS} search.

\begin{figure*}[htbp]
\includegraphics[width=\textwidth]{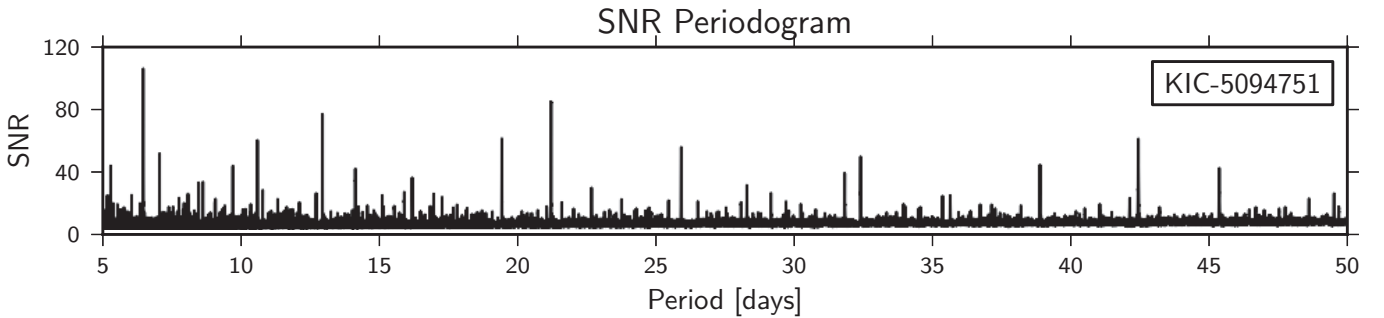}
\caption{SNR periodogram of KIC-5094751 photometry, demonstrating \TERRA's insensitivity to lower SNR candidates in multi-candidate systems. \cite{Batalha12} lists two planets belonging to KIC-5094751, KOI-123.01 and KOI-123.02 with \Per = 6.48 and 21.22 days, respectively. \TERRA detected KOI-123.01 with a period of 6.48 days (highest SNR peak). Sub-harmonics belonging to KOI-123.01 are visible at $[2, 3, \hdots] \times \Per = [13.0, 19.4, \hdots]$~days. A second set of SNR peaks due to KOI-123.02 (\Per = 21.2 days) is visible at $[0.5, 2, \hdots] \times \Per = [10.6, 42.4, \hdots]$~days. Had we removed the transit due to KOI-123.01, KOI-123.02 would be easily detectible due its high SNR of $\sim80$. \TERRA does not yet include multi-candidate logic and is thus blind to lower SNR candidates in multi-candidate systems.}
\label{fig:SNRmulti}
\end{figure*}

Since \TERRA only passes the (\Per,\ep,\tdur) triple with the highest SNR on to DV, \TERRA does not detect additional planets with lower SNR due to either smaller size or longer orbital period. As an example of \TERRA's insensitivity to small candidates in multi-candidate systems, we show the \TERRA SNR periodogram for KIC-5094751 in Figure~\ref{fig:SNRmulti}. \cite{Batalha12} lists two candidates belonging to KIC-5094751: KOI-123.01 and KOI-123.02 with P = 6.48 and 21.22 days, respectively. Although the SNR periodogram shows two sets of peaks coming from two distinct candidates, \TERRA only identifies the first peak.  Automated identification of multi-candidate systems is a planned upgrade for \TERRA. Another caveat is that \TERRA assumes strict periodicity and struggles to detect low SNR transits with significant transit timing variations, i.e. variations longer than the transit duration.

\subsection{Data Validation}
\label{sec:TERRAdv}
If the SNR periodogram has a maximum SNR peak > 12, we flag the corresponding (\Per,\ep,\tdur) for additional vetting. Following the language of the official \Kepler pipeline, we refer to these triples as ``threshold crossing events'' (TCEs), since they have high photometric SNR, but are not necessarily consistent with an exoplanet transit. \TERRA vets the TCEs in a step called ``data validation,'' again following the nomenclature of the official \Kepler pipeline. Data validation (DV), as implemented in the official \Kepler pipeline, is described in \cite{Jenkins10}. We emphasize that \TERRA DV does not depend on the DV component of the \Kepler team pipeline.

\begin{figure}
\begin{center}
\includegraphics[width=0.48\textwidth]{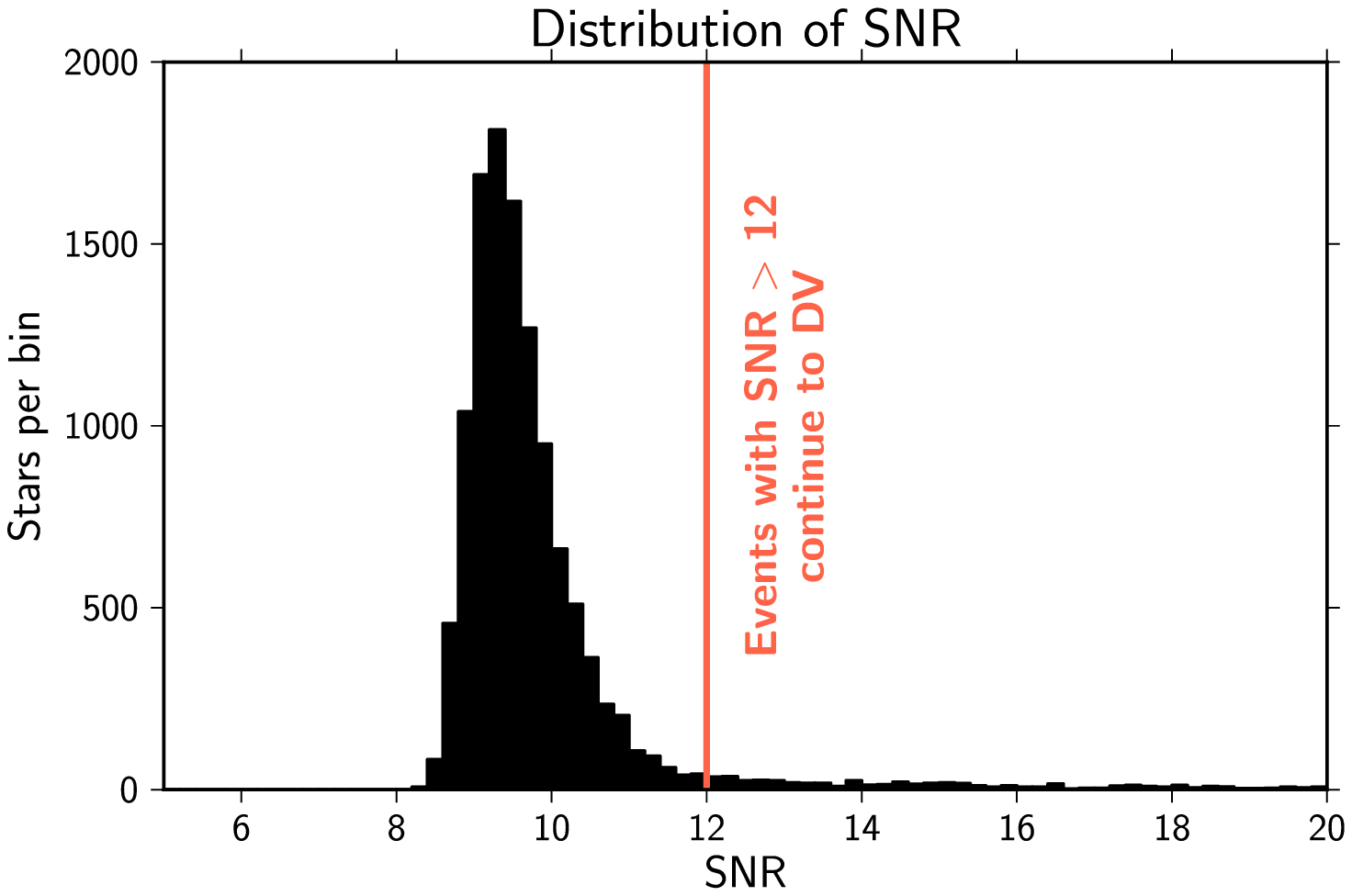}
\caption{Distribution of the highest SNR peak for each star in the Best12k sample. We show SNR = 5--20 to highlight the distribution of low SNR events. The \NSNR stars with SNR > 12 are labeled ``threshold crossing events'' (TCEs) and are subjected to additional scrutiny in the ``data validation'' component of \TERRA.}
\label{fig:SNR-hist}
\end{center}
\end{figure}

We show the distribution of maximum SNR for each Best12k star in Figure~\ref{fig:SNR-hist}. Among the Best12k stars, \NSNR have a maximum SNR peak exceeding 12. Adopting SNR = 12 as our threshold balances two competing needs: the desire to recover small planets (low SNR) and the desire to remove as many non-transit events as possible  before DV (high SNR). As discussed below, only 129 out of all 738 events with SNR > 12 are consistent with an exoplanet transit, with noise being responsible for the remaining \NSNRnoteKOI. As shown in Figure~\ref{fig:SNR-hist}, that number grows rapidly as we lower the SNR threshold. For example, the number of TCEs grows to 3055 with a SNR threshold of 10, dramatically increasing the burden on the DV component.

A substantial number (\NSNRnotP) of TCEs are due to harmonics or subharmonics of TCEs outside of the \Per= 5--50 day range and are discarded. In order to pass DV, a TCE must also pass a suite of four diagnostic metrics. The metrics are designed to test whether a light curve is consistent with an exoplanet transit. We describe the four metrics in Table~\ref{tab:DV} along with the criteria the TCE must satisfy in order pass DV. The metrics and cuts were determined by trial and error. We recognize that the \TERRA DV metrics and cuts are not optimal and discard a small number of compelling exoplanet candidates, as discussed in Section~\ref{sec:catonly}. However, since we measure \TERRA's completeness by injection and recovery of synthetic transits, the sub-optimal nature of our metrics and cuts is incorporated into our completeness corrections. 

\begin{figure*}
\begin{center}
\includegraphics[width=\textwidth]{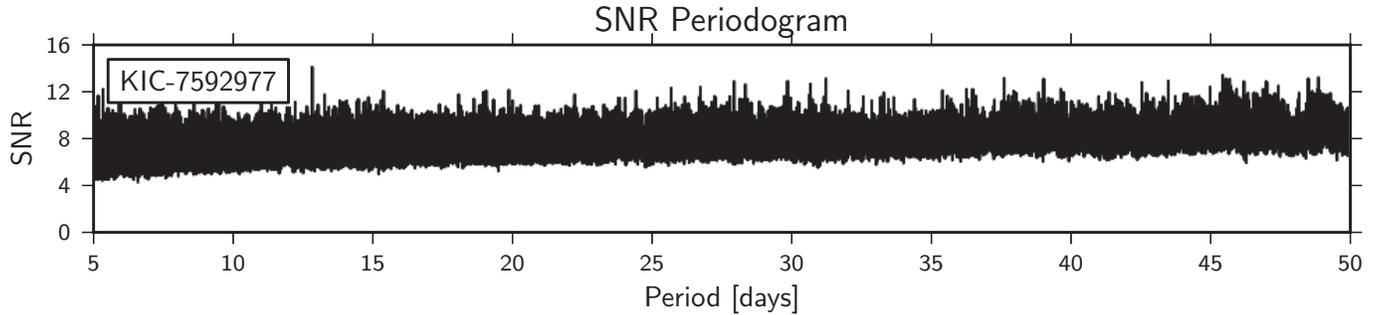}
\caption{SNR periodogram of KIC-7592977, which passed the automated DV cuts, but was removed manually. KIC-7592977 photometry exhibited short bursts of high photometric scatter, which raised the noise floor to SNR $\sim$ 12, up from SNR $\sim$ 8 as in most stars.}
\label{fig:SNR-bad}
\end{center}
\end{figure*}

Our suite of automated cuts removes all but 145 TCEs. We perform a final round of manual vetting and remove 16 additional TCEs, leaving 129 planet candidates. Most TCEs that we remove manually come from stars with highly non-stationary photometric noise properties. Some stars have small regions of photometry that exceed typical noise levels by a factor of 3. We show the SNR periodogram for one such star, KIC-7592977, in Figure~\ref{fig:SNR-bad}. Our definition of SNR (Equation~\ref{eqn:SNR}) incorporates a single measure of photometric scatter based on the median absolute deviation, which is insensitive to short bursts of high photometric variability. In such stars, fluctuations readily produce SNR $\sim$ 12 events and raise the detectability floor to SNR $\sim$ 12, up from SNR $\sim$ 8 in most stars. We also visually inspect phase-folded light curves for coherent out-of-transit variability, not caught by our automated cuts, and for evidence of a secondary eclipse.

\begin{deluxetable*}{  l p{4.5in} r }
\tablecaption{In order to pass the ``data validation'' (DV) stage, a ``threshold crossing event'' (TCE) must pass the following suite of cuts.}
\tablewidth{0pt}
\tablehead{\colhead{name} &	\colhead{description} & \colhead{value}}
\startdata
     {\tt s2n\_out\_on\_in }
     & 
     Compelling transits have flat out-of-transit light curves. For a TCE with (\Per,\ep,\tdur), we remove the transit region from the light curve and evaluate the SNR of all other ($P',\ep',\tdur'$) triples where $P=P'$ and $\tdur=\tdur'$. {\tt s2n\_out\_on\_in} is the ratio of the two highest SNR events.
     &
     < 0.7
     \\
     {\tt med\_on\_mean }
     & 
     Since the our definition of SNR (Equation~\ref{eqn:SNR}) depends on the arithmetic mean of individual transit depths, outliers occasionally produce   
high SNR TCEs. For each TCE, we compute a robust SNR,
    \[
	\text{medSNR} = \frac{\sqrt{N_{T}}}{\sigma} \text{median}(\mdf_{i}).
    \]
     {\tt med\_on\_mean} is medSNR divided by SNR as defined in Equation~\ref{eqn:SNR}.        &
     > 1.0
     \\
     {\tt autor }
     &
     We compute the circular autocorrelation of the phase-folded light curve. {\tt autor} is the ratio of the highest autocorrelation peak (at 0 lag) to the second highest peak and is sensitive to out-of-transit variability.    
     &
     > 1.6
	\\
	{\tt taur}
	& 
	We fit the phase-folded light curve with a \cite{Mandel02} model. {\tt taur} is the ratio of the best fit transit duration to the maximum duration given the KIC stellar parameters and assuming a circular orbit.
	&
	< 2.0
	\\
\enddata
\label{tab:DV}
\end{deluxetable*}

\section{Small Planets Found by \TERRA}
\label{sec:TERRAplanetYield}
Out of the 12,000 stars in the Best12k sample, \TERRA detected 129 planet candidates achieving SNR > 12 that passed our suite of DV cuts as well as visual inspection. Table~\ref{tab:planetsTERRA} lists the 129 planet candidates. We derive planet radii using \Rp/\Rstar (from Mandel-Agol model fits) and \Rstar from spectroscopy (when available) or broadband photometry. 

We obtained spectra for 100 of the 129 stars using HIRES (Vogt et al. 1994) at the Keck I telescope with standard configuration of the California Planet Survey \citep{Marcy08}. These spectra have resolution of $\sim$ 50,000, at a signal-to-noise of 45 per pixel at 5500~\AA. We determine stellar parameters using a routine called {\tt \SpecMatch} (Howard et al. 2013, in prep). In brief, {\tt \SpecMatch} compares a stellar spectrum to a library of $\sim800$ spectra with \teff = 3500--7500~K and \logg = 2.0--5.0 (determined from LTE spectral modeling). Once the target spectrum and library spectrum are placed on the same wavelength scale, we compute $\chi^2$, the sum of the squares of the pixel-by-pixel differences in normalized intensity. The weighted mean of the ten spectra with the lowest $\chi^2$ values is taken as the final value for the effective temperature, stellar surface gravity, and metallicity. We estimate {\tt \SpecMatch}-derived stellar radii are uncertain to 10\% RMS, based on tests of stars having known radii from high resolution spectroscopy and asteroseismology.

For 27 stars where spectra are not available, we adopt the photometrically-derived stellar parameters of \cite{Batalha12}. These parameters are taken from the KIC \citep{Brown11}, but then modified so that they lie on the Yonsei-Yale stellar evolution models of \cite{Demarque04}. The resulting stellar radii have uncertainties of 35\% (rms), but can be incorrect by a factor of 2 or more. As an extreme example, the interpretations of the three planets in the KOI-961 system \citep{Muirhead12} changed dramatically when HIRES spectra showed the star to be an M5 dwarf (0.2~\Rsun as opposed to 0.6~\Rsun listed in the KIC). We could not obtain spectra for two stars, KIC-7345248 and KIC-8429668, which were not present in \cite{Batalha12}. We determine stellar parameters for these stars by fitting the KIC photometry to Yonsei-Yale stellar models. We adopt 35\% fractional errors on photometrically-derived stellar radii.

Once we determine \Per and \ep, we fit a \cite{Mandel02} model to the phase-folded photometry. Such a model has three free parameters: \rrat, the planet to stellar radius ratio; $\tau$, the time for the planet to travel a distance \Rstar during transit; and $b$, the impact parameter. In this work, \rrat is the parameter of interest. However, $b$ and \rrat are covariant, i.e. a transit with $b$ approaching unity only traverses the limb of the star, and thus produces a shallower transit depth. In order to account for this covariance, best fit parameters were computed via Markov Chain Monte Carlo. We find that the fractional uncertainty on \rrat, $\frac{\sigma(\rrat)}{\rrat}$ can be as high as 10\%, but is generally less than 5\%. Therefore, the error on \Rp due to covariance with $b$ is secondary to the uncertainty on \Rstar.

\begin{figure*}
\centering
\includegraphics[width=1.5\columnwidth]{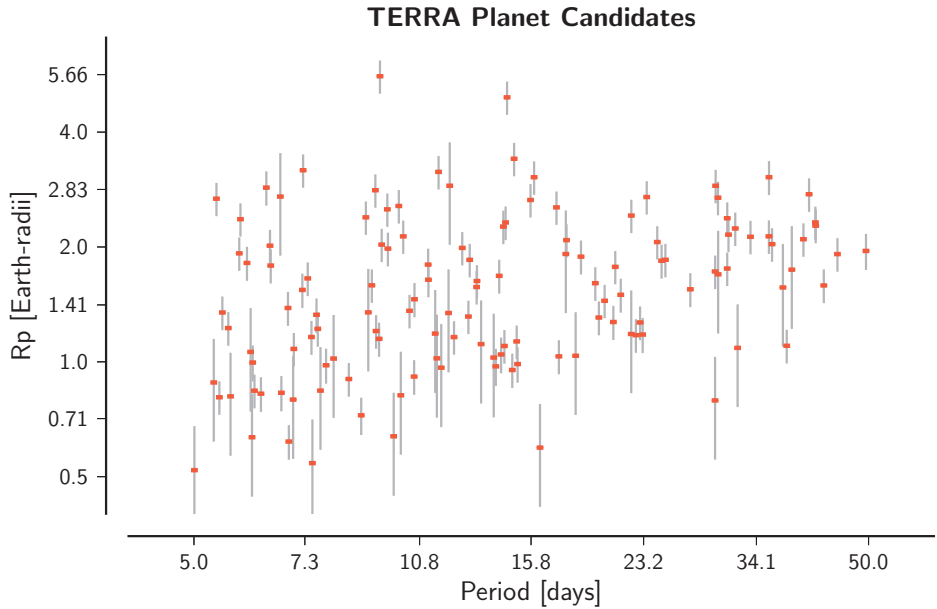}
\caption{Periods and radii of 129 planet candidates detected by \TERRA. Errors on \Rp are computed via 
$\frac{ \sigma(\Rp) }{\Rp} =  
\sqrt{
\left( \frac{\sigma(\Rstar)}{\Rstar} \right)^2 +
\left( \frac{\sigma(\rrat)}{\rrat}\right)^2
}$, where \rrat is the radius ratio. The error in \Rp stems largely from the uncertainty in stellar radii. We adopt $\frac{\sigma(\Rstar)}{\Rstar}$ = 10\% for the 100 stars with spectroscopically determined \Rstar and $\frac{\sigma(\Rstar)}{\Rstar}$ = 35\% for the remaining stars with \Rstar determined from photometry. Using MCMC, we find the uncertainty in \rrat  is generally < 5\% and thus a minor component of the overall error budget.}
\label{fig:TERRAraw}
\end{figure*}

\begin{figure*}
\centering
\includegraphics[width=1.5\columnwidth]{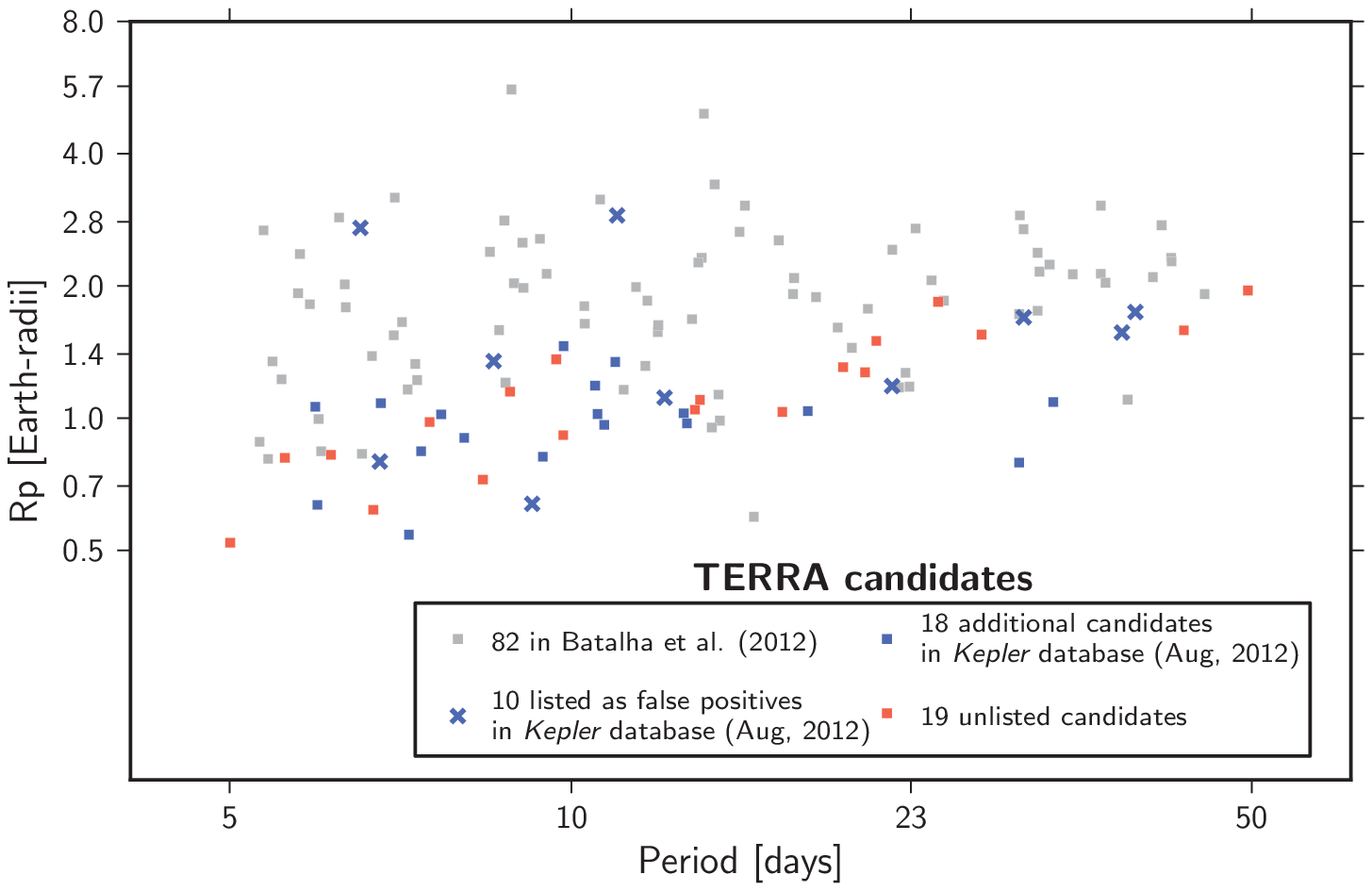}
\caption{Periods and radii of all 129 \TERRA planet candidates. The gray points show candidates that were listed in \cite{Batalha12}. The blue crosses represent candidates deemed false positives by the \Kepler team as of August 8, 2012 (Jason Rowe, private communication 2012). These false positives are removed from our sample prior to computing occurrence. Eighteen additional candidates were listed in the same \Kepler team database. Red points show 19 unlisted \TERRA candidates.}
\label{fig:TERRAcommon}
\end{figure*}

We show the distribution of \TERRA candidates in Figure~\ref{fig:TERRAraw} over the two-dimensional domain of planet radius and orbital period. Our 129 candidates range in size from 6.83 \Re to 0.48 \Re (smaller than Mars). The median \TERRA candidate size is 1.58 \Re. In Figure~\ref{fig:TERRAcommon}, we show the substantial overlap between the \TERRA planet sample and those produced by the \Kepler team. \TERRA recovers 82 candidates listed in \cite{Batalha12}. We discuss the significant overlap between the two works in detail in Section~\ref{sec:compare}. As of August 8th, 2012, 10 of our \TERRA candidates were listed as false positives in an internal database of \Kepler planet candidates maintained by Jason Rowe (Jason Rowe, 2012, private communication) and are shown as blue crosses in Figure~\ref{fig:TERRAcommon}. We do not include these 10 candidates in our subsequent calculation of occurrence. Table~\ref{tab:planetsTERRA} lists the KIC identifier, best fit transit parameters, stellar parameters, planet radius, and \Kepler team false positive designation of all 129 candidates revealed by the \TERRA algorithm. The best fit transit parameters include orbital period, \Per; time of transit center, \ep; planet to star radius ratio, \rrat; time for planet to cross \Rstar during transit, $\tau$; and impact parameter, $b$. We list the following stellar properties: effective temperature, \teff; surface gravity, \logg; and stellar radius, \Rstar.

\section{Completeness of Planet Catalog}
\label{sec:MC}
When measuring the distribution of planets as a function of \Per and \Rp, understanding the number of missed planets is as important as finding planets themselves. H12 accounted for completeness in a rough sense based on signal-to-noise considerations. For each star in their sample, they estimated the SNR over a range of \Per and \Rp using CDPP as an estimate of the photometric noise on transit-length timescales. H12 chose to accept only planets with SNR > 10 in a single quarter of photometry for stars brighter than Kp = 15.  This metric used CDPP and was a reasonable pass on the data, particularly when the pipeline completeness was unknown.  Determining expected SNR from CDPP does not incorporate the real noise characteristics of the photometry, but instead approximates noise on transit timescales as stationary (CDDP assumed to be constant over a quarter) and Gaussian distributed. Moreover, identifying small transiting planets with transit depths comparable to the noise requires a complex, multistage pipeline. Even if the integrated SNR is above some nominal threshold, the possibility of missed planets remains a concern.

We characterize the completeness of our pipeline by performing an extensive suite of injection and recovery experiments. We inject mock transits into raw photometry, run this photometry though the same pipeline used to detect planets, and measure the recovery rate. This simple, albeit brute force, technique captures the idiosyncrasies of the \TERRA pipeline that are missed by simple signal-to-noise considerations.

We perform 10,000 injection and recovery experiments using the following steps:

\begin{enumerate}
\item We select a star randomly from the Best12k sample. 
\item We draw (\Per,\Rp) randomly from log-uniform distributions over 5--50~days and 0.5--16.0~\Re.
\item We draw impact parameter and orbital phase randomly from uniform distributions ranging from 0 to 1.
\item We generate a \cite{Mandel02} model.
\item We inject it into the ``simple aperture photometry'' of the selected star.
\end{enumerate}
We then run the calibration, grid-based search, and data validation components of \TERRA (Sections~\ref{sec:TERRAcal}, \ref{sec:TERRAgrid}, and \ref{sec:TERRAdv}) on this photometry and calculate the planet recovery rate. We do not, however, perform the visual inspection described in Section~\ref{sec:TERRAdv}. An injected transit is considered recovered if the following two criteria are met: (1) The highest SNR peak passes all DV cuts and (2) the output period and epoch are consistent with the injected period and epoch to within 0.01 and 0.1 days, respectively.

\begin{figure*}[htbp]
\includegraphics[width=\textwidth]{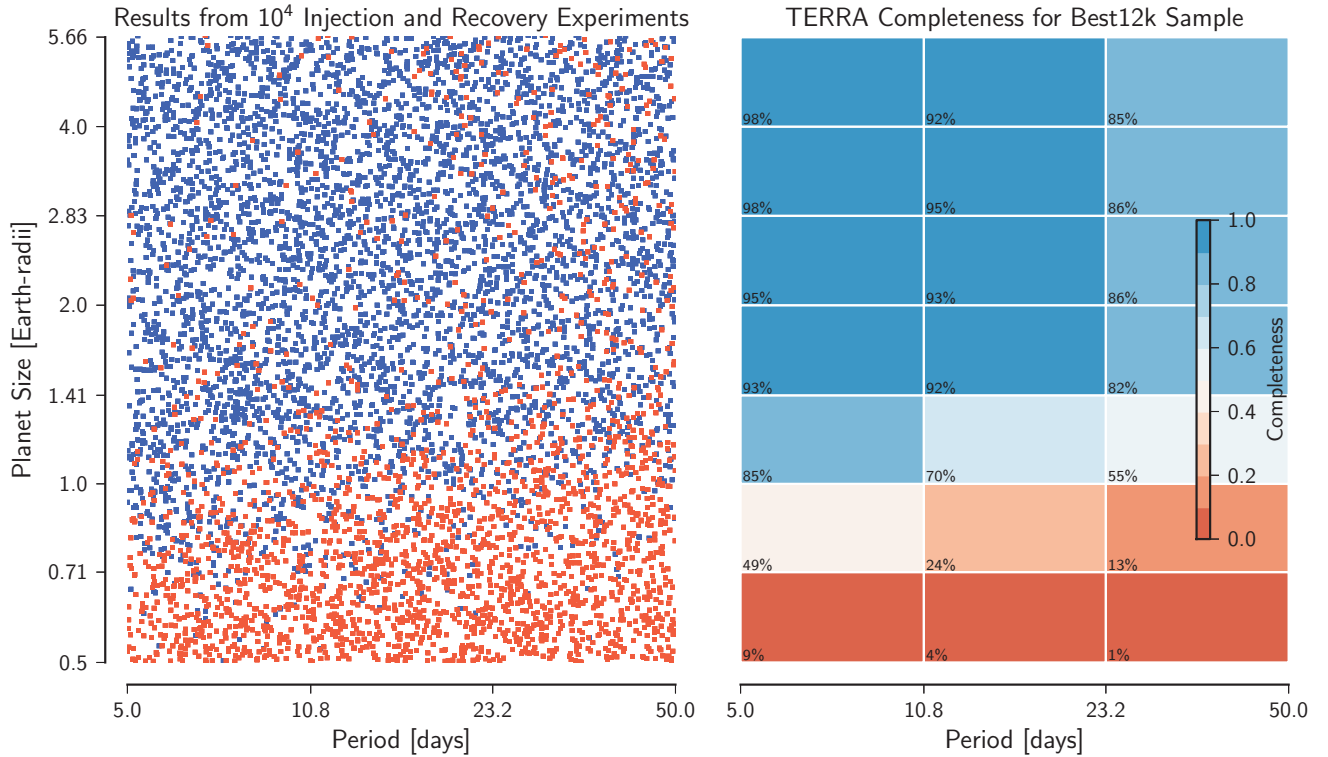}
\caption{Results from the injection and recovery of 10,000 synthetic transit signals into actual photometry of randomly selected stars from our Best12k stellar sample. Each point represents the planet radius and orbital period of a mock transiting planet. The blue points represent signals that passed the DV post-analysis and where \TERRA recovers the correct period and epoch. Signals that did not pass DV and/or were not successfully recovered, are shown as red points. Pipeline completeness is simply the number of blue points divided by the total number of points in each bin. The figure shows that for planet sizes above 1.0~\Re, our pipeline discovers over 50\% of the injected planets, and presumably accomplishes a similar success rate for actual transiting planets. The completeness for planets larger than 1~\Re is thus high enough to compute planet occurrence for such small planets, with only moderate completeness corrections needed (less than a factor of 2). Note that we are measuring the recovery rate of putative planets in the Best12k sample with \TERRA. Had we selected a lower noise stellar sample, for example the ``Best6k,'' the region of high completeness would extend to even small radii.} 
\label{fig:comp}
\end{figure*}

Figure~\ref{fig:comp} shows the distribution of recovered simulations as a function of period and radius. Nearly all simulated planets with \Rp > 1.4 \Re are recovered, compared to almost none with \Rp < 0.7 \Re. Pipeline completeness is determined in small bins in (\Per,\Rp)-space by dividing the number of successfully recovered transits by the total number of injected transits in a bin-by-bin basis. This ratio is \TERRA's recovery rate of putative planets within the Best12k sample. Thus, our quoted completeness estimates only pertain to the low photometric noise Best12k sample. Had we selected an even more rarified sample, e.g. the ``Best6k,'' the region of high completeness would extend down toward smaller planets. 

\section{Occurrence of Small Planets}
\label{sec:occurrenceme}

Following H12, we define planet occurrence, $f$, as the fraction of a defined population of stars having planets within a domain of planet radius and period, including all orbital inclinations. \TERRA, however, is only sensitive to one candidate (highest SNR) per system, so we report occurrence as the fraction of stars with {\em one or more} planets with \Per = 5--50 days. Our occurrence measurements apply to the Best12k sample of low-noise, solar-type stars described in Section~\ref{sec:Sample}. 

In computing planet occurrence in the Best12k sample, we follow the prescription in H12 with minor modifications. Notably, we have accurate measures of detection completeness described in the previous section. In contrast, H12 estimated completeness based on the presumed signal-to-noise of the transit signal, suffering both from approximate characterization of photometric noise using CDPP and from poor knowledge of the efficiency of the planet-finding algorithm for all periods and sizes. 

For each \Per-\Rp bin, we count the number of planet candidates, \Np. Each planet that transits represents many that do not transit given the orientation of their orbital planes with respect to \Kepler's line of sight. Assuming random orbital alignment, each observed planet represents a/\Rstar total planets when non-transiting geometries are considered. For each cell, we compute the number of augmented planets, $\NpAug = \sum_i a_i/R_{\star,i}$, which accounts for planets with non-transiting geometries. We then use Kepler's \nth{3} law together with \Per and $M_\star$ to compute a/\Rstar assuming a circular orbit.\footnote{H12 determined a/\Rstar directly from light curve fits, but found little change when computing occurrence from a/\Rstar using Kepler's third law.}

\begin{figure*}
\centering
\includegraphics[width=\textwidth]{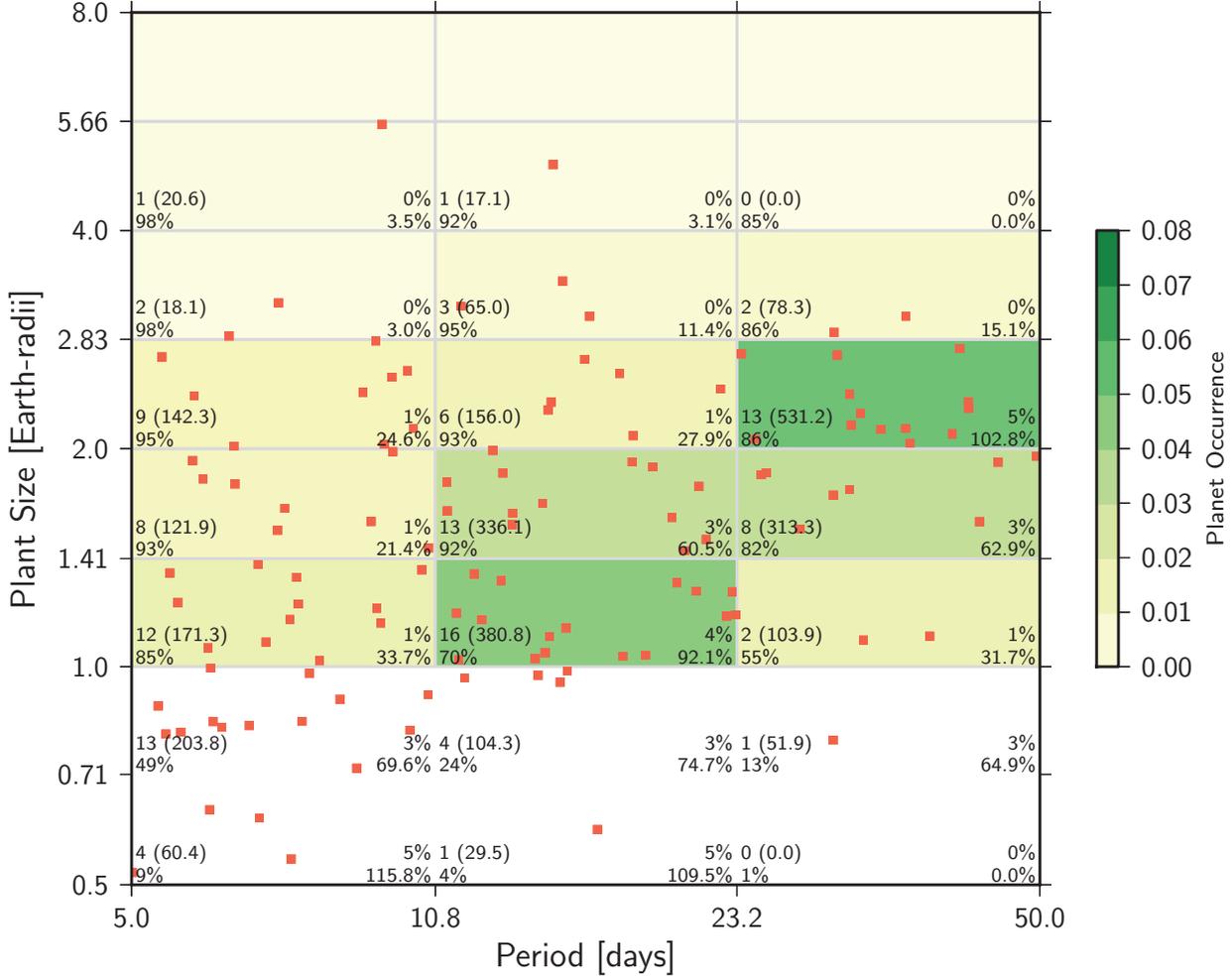}
\caption{Planet occurrence as a function of orbital period and planet radius for \Per = 5--50 days and \Rp = 0.5--8 \Re. \TERRA planet candidates are shown as red points. Cell occurrence, \fcell, is given by the color scale. We quote the following information for each cell: Top left--number of planets (number of augmented planets); lower left--completeness; top right--fractional planet occurrence, \fcell; bottom right--normalized planet occurrence, \flogA. We do not color cells where the completeness is less than 50\% (i.e. the completeness correction is larger than a factor of 2).}
\label{fig:TERRAoccur2D}
\end{figure*}

To compute occurrence, we divide the number of stars with planets in a particular cell by the number of stars amenable to the detection of a planet in a given cell, $\NstarAmen$. This number is just $N_\star$ = 12,000 times the completeness, computed in our Monte Carlo study. The debiased fraction of stars with planets per \Per-\Rp bin, $\fcell$, is given by $\fcell=\NpAug/\NstarAmen$. We show \fcell on the \Per-\Rp plane in Figure~\ref{fig:TERRAoccur2D} as a color scale. We also compute \flogA, i.e. planet occurrence divided by the logarithmic area of each cell, which is a measure of occurrence which does not depend on bin size. We annotate each \Per-\Rp bin of Figure~\ref{fig:TERRAoccur2D} with the corresponding value of \Np, \NpAug, \fcell, and \flogA.

Due to the small number of planets in each cell, errors due to counting statistics alone are significant. We compute Poisson errors on \Np for each cell. Errors on \NpAug, \fcell, and \flogA include only the Poisson errors from \Np. There is also shot noise associated with the Monte Carlo completeness correction due to the finite number of simulated planets in each \Per-\Rp cell, but such errors are small compared to errors on \Np. The orbital alignment correction, $a/\Rstar$, is also uncertain due to imperfect knowledge of stellar radii and orbital separations. We do not include such errors in our occurrence estimates.

\begin{figure}
\includegraphics[width=\columnwidth]{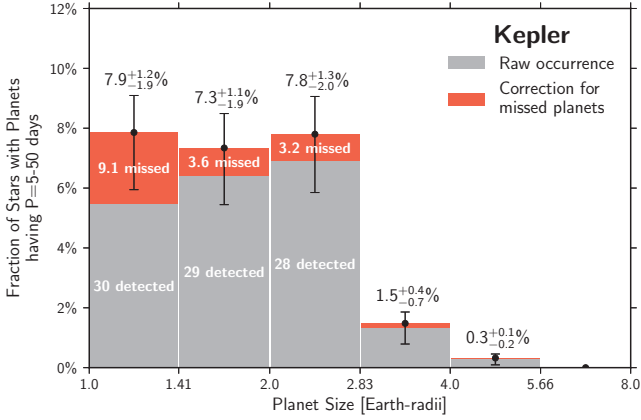}
\caption{Distribution of planet occurrence for \Rp ranging from 1.0 to 8.0~\Re. We quote the fraction of Sun-like stars harboring a planet with \Per~=~5--50~days for each \Rp bin. We observe a rapid rise in planet occurrence from 8.0 down to 2.8~\Re, as seen in H12. Below 2.8 \Re, the occurrence distribution is consistent with flat. This result rules out a power law increase in planet occurrence toward smaller radii. Adding up the two smallest radius bins, we find  \fonetwo of Sun-like stars harbor a 1.0--2.0~\Re planet within $\sim0.25$~AU. To compute occurrence as a function of \Rp, we simply sum occurrence rates for all period bins shown in Figure~\ref{fig:TERRAoccur2D}.  Errors due to counting statistics are computed by adding errors from each of the three period bins in quadrature. The gray portion of the histogram shows occurrence values before correcting for missed planets due to pipeline incompleteness. Our correction to account for missed planets is shown in red, and is determined by the injection and recovery of synthetic transits described in Section~\ref{sec:MC}. We do not show occurrence values where the completeness is < 50\%.}
\label{fig:TERRAoccur1D}
\end{figure}

Of particular interest is the distribution of planet occurrence with \Rp for all periods. We marginalize over \Per by summing occurrence over all period bins from 5 to 50~days. The distribution of radii shown in Figure~\ref{fig:TERRAoccur1D} shows a rapid rise in occurrence from 8.0 to 2.8~\Re. H12 also observed a rising occurrence of planets down to 2.0~\Re, which they modeled as a power law. {\em Planet occurrence is consistent with a flat distribution from 2.8 to 1.0~\Re, ruling out a continuation of a power law increase in occurrence for planets smaller than 2.0~\Re. We find \fonetwo of Sun-like stars harbor a 1.0--2.0~\Re planet with \Per~=~5--50~days.} Including larger planets, we find that $24.8^{+2.1}_{-3.4}\%$ of stars harbor a planet larger than Earth with \Per~=~5--50~days. Occurrence values assuming a 100\% efficient pipeline are shown as gray bars in Figure~\ref{fig:TERRAoccur1D}. The red bars show the magnitude of our completeness correction. Even though \TERRA detects many planets smaller than 1.0~\Re, we do not report occurrence for planets smaller than Earth since pipeline completeness drops abruptly below 50\%.

\begin{figure}
\includegraphics[clip=True,width=1\columnwidth]{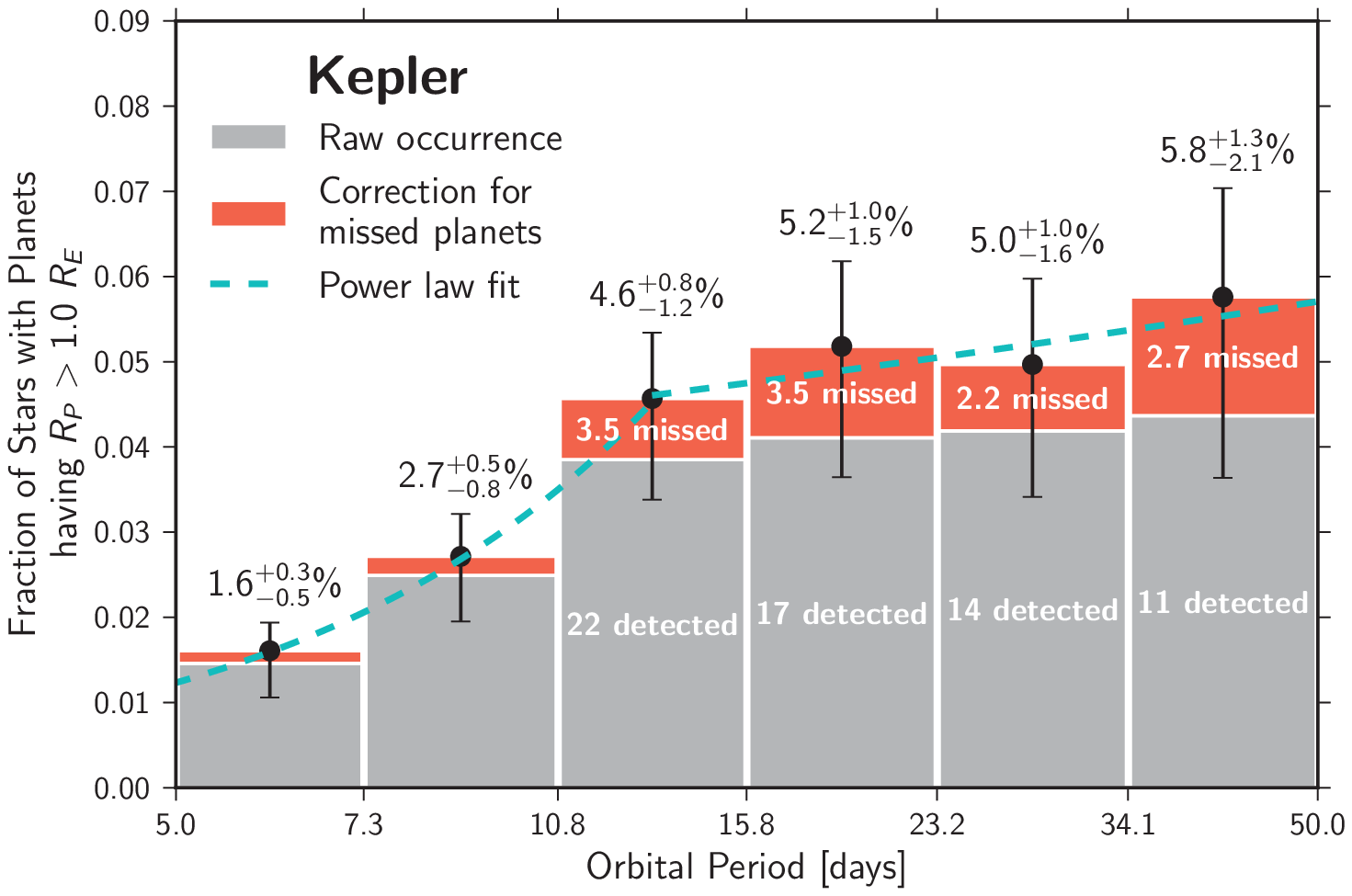}
\caption{Distribution of planet occurrence for different orbital periods ranging from 5 to 50~days. We quote the fraction of Sun-like stars with a planet Earth-size or larger as a function of orbital period. We observe a gradual rise in occurrence from 5.0 to 10.8~\Re followed by a leveling off for longer orbital periods. H12 observed a similar leveling off in their analysis which included planets larger than 2~\Re. We fit the domains above and below 10.8~\Re separately with power laws, $d f / d \log \Per = k_{P} \Per^{\alpha}$. We find best fit values of $k_{P} = 0.185^{+0.043}_{-0.035},\ \alpha =0.16 \pm 0.07$ for \Per~=~5--10.8~days and $k_{P} = 8.4^{+0.9}_{-0.8} \times 10^{-3},\ \alpha =1.35 \pm 0.05$ for \Per~=~10.8--50~days. Speculatively, we extrapolate the latter power law fit another decade in period and estimate $41.7^{+6.8}_{-5.9}$\% of Sun-like stars harbor a planet Earth-size or larger with \Per~=~50--500~days. As in Figure~\ref{fig:TERRAoccur1D}, the gray portion of the histogram shows uncorrected occurrence while the red region shows our correction for pipeline incompleteness. Note that the number of detected planets decreases as \Per increases from 10.8 to 50~days, while occurrence remains nearly constant. At longer periods, the geometric transit probability is lower, and each detected planet counts more toward $d f / d \log \Per$.}
\label{fig:TERRAoccurP}
\end{figure}

We show planet occurrence as a function of orbital period in Figure~\ref{fig:TERRAoccurP}. In computing this second marginal distribution, we include radii larger than 1~\Re so that corrections due to incompleteness are small. Again, as in Figure~\ref{fig:TERRAoccur1D}, gray bars represent uncorrected occurrence values while red bars show our correction to account for planets that \TERRA missed. Planet occurrence rises as orbital period increases from 5.0 to 10.8~days. Above 10.8~days, planet occurrence is nearly constant per logarithmic period bin with a slight indication of a continued rise. This leveling off of the distribution was noted by H12, who considered \Rp > 2.0 \Re. We fit the distribution of orbital periods for \Rp > 1.0 \Re with two power laws of the form
\begin{equation}
\frac{d f }{ d \log \Per} = k_{P} \Per^{\alpha}, 
\end{equation}
where $\alpha$ and $k_{P}$ are free parameters. We find best fit values of $k_{P} = 0.185^{+0.043}_{-0.035},\ \alpha =0.16 \pm 0.07$ for \Per~=~5--10.8~days and $k_{P} = 8.4^{+0.9}_{-0.8} \times 10^{-3},\ \alpha =1.35 \pm 0.05$ for \Per~=~10.8--50~days. We note that $k_{P}$ and $\alpha$ are strongly covariant. Extrapolating the latter fit speculatively to \Per > 50~days, we find $41.7^{+6.8}_{-5.9}$\% of Sun-like stars host a planet 1 \Re or larger with \Per~=~50--500~days.

\section{Comparison of \TERRA and Batalha et al. (2012) Planet Catalogs}
\label{sec:compare}
Here, we compare our candidates to those of \cite{Batalha12}. Candidates were deemed in common if their periods agree to within 0.01 days. We list the union of the \TERRA and \cite{Batalha12} catalogs in Table~\ref{tab:planetsJOIN}. Eighty-two candidates appear in both catalogs (Section~\ref{ssec:common}), 47 appear in this work only (Section~\ref{ssec:TERRAOnly}), and 33 appear in \cite{Batalha12} only (Section~\ref{sec:catonly}). We discuss the significant overlap between the two catalogs and explain why some candidates were detected by one pipeline but not the other.

\subsection{Candidates in Common}
\label{ssec:common}
Eighty-two of our candidates appear in the \cite{Batalha12} catalog. We show these candidates in \Per-\Rp space in Figure~\ref{fig:TERRAcommon} as grey points. \TERRA detected no new candidates with \Rp > 2 \Re. This agreement in detected planets having \Rp > 2 \Re demonstrates high completeness for such planets in both pipelines for this sample of quiet stars. This is not very surprising since candidates with \Rp > 2 \Re have high SNR, e.g. min, median, and max SNR = 19.3, 71.5, and 435 respectively.

Radii for the 82 planets in common were fairly consistent between \cite{Batalha12} and this work. The two exceptions were KIC-8242434 and KIC-8631504. Using {\tt SpecMatch}, we find stellar radii of 0.68 and 0.72~\Rsun, respectively, down from 1.86 and 1.80 \Rsun  in \cite{Batalha12}. The revised planet radii are smaller by over a factor of two. Radii for the other planets in common were consistent to $\sim$ 20\%.

\subsection{\TERRA Candidates Not in \cite{Batalha12} Catalog}
\label{ssec:TERRAOnly}

\TERRA revealed 47 planet candidates that did not appear in \cite{Batalha12}. Such candidates are colored blue and red in Figure~\ref{fig:TERRAcommon}. Many of these new detections likely stem from the fact that we use twice the photometry that was available to Batalha et al. (2012). To get a sense of how additional photometry improves the planet yield of the \Kepler pipeline beyond \cite{Batalha12}, we compared the \TERRA candidates to the \Kepler team KOI list dated August 8, 2012 (Jason Rowe, private communication). The 28 candidates in common between the August 8, 2012 \Kepler team sample and this work are colored blue in Figure~\ref{fig:TERRAcommon}. Of these 28 candidates, 10 are listed as false positives and denoted as crosses in Figure~\ref{fig:TERRAcommon}.

We announce 37 new planet candidates with respect to \cite{Batalha12} that were not listed as false positives in the \Kepler team sample. These 37 candidates, all with \Rp~$\lessapprox$ 2~\Re, are a subset of those listed in Table~\ref{tab:planetsTERRA}. As a convenience, we show this subset in Table~\ref{tab:planetsTERRAnew}. We remind the reader that all photometry used in this work is publicly available. We hope that interested readers will fold the photometry on the ephemeris in Table~\ref{tab:planetsTERRA} and access critically whether a planet interpretation is correct. As a quick reference, we have included plots of the transits of the 37 new candidates from Table~\ref{tab:planetsTERRAnew} in the appendix (Figures~\ref{fig:binned1} and \ref{fig:binned2}). We do not claim that our additional candidates bring pipeline completeness to unity for planets with \Rp~$\lessapprox$ 2~\Re. As shown in Section~\ref{sec:MC}, our planet sample suffers from significant incompleteness in the same \Per-\Rp space where most of the new candidates emerged.

\subsection{\cite{Batalha12} Candidates Not in \TERRA catalog}
\label{sec:catonly}
\begin{figure}[htbp]
\includegraphics[width=\columnwidth]{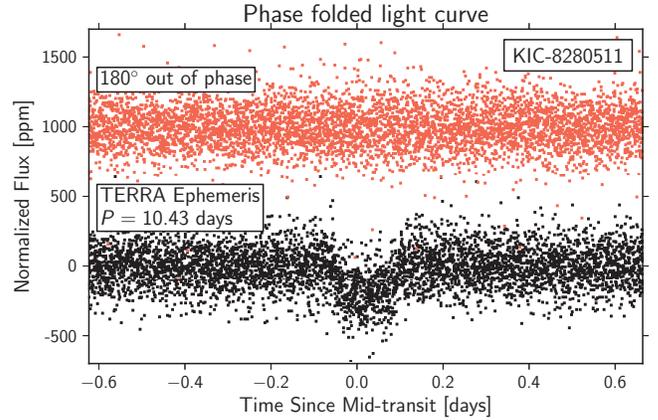}
\caption{Phase-folded photometry of KIC-8280511 folded on the correct 10.43~day period found by \TERRA. KOI-1151.01 is listed with \Per = 5.22~days in Batalha et al. (2012). If the transit was truly on the 5.22~day period, we should see a transit of equal depth 180 degrees out of phase. KOI-1151.01 is listed in \cite{Batalha12} with half its true period.}
\label{fig:KIC-8280511}
\end{figure}

There are 33 planet candidates in the \cite{Batalha12} catalog from Best12k stars that \TERRA missed. Of these, 28 are multi-candidate systems where one component was identified by \TERRA. \TERRA is currently insensitive to multiple planet systems (as described in Section~\ref{sec:TERRAgrid}). \TERRA missed the remaining 5 \cite{Batalha12} candidates for the following reasons:
\begin{itemize}
\item 2581.01 : A bug in the pipeline prevented successful photometric calibration (Section~\ref{sec:TERRAcal}). This bug affected 19 out of   12,000 stars in the Best12k sample.
\item 70.01, 111.01, 119.01 : Failed one of the automated DV cuts ({\tt taur}, {\tt med\_on\_mean}, and {\tt taur}, respectively). We examined these three light curves in the fashion described in Section~\ref{sec:TERRAdv}, and we determined these light curves were consistent with an exoplanet transit. The fact that DV is discarding compelling transit signals decreases \TERRA's overall completeness. Computing DV metrics and choosing the optimum cuts is an art. There is room for improvement here.
\item KOI-1151.01 : Period misidentified in \cite{Batalha12}. In \cite{Batalha12} KOI-1151.01 is listed with with a \Per~=~5.22~days. \TERRA found a candidate with \Per~=~10.43~days. Figure~\ref{fig:KIC-8280511} shows phase-folded photometry with the \TERRA ephemeris. A period of 5.22 days would imply dimmings in regions where the light curve is flat.
\end{itemize}

\begin{figure*}[htbp]
\centering
\includegraphics[width=1.5\columnwidth]{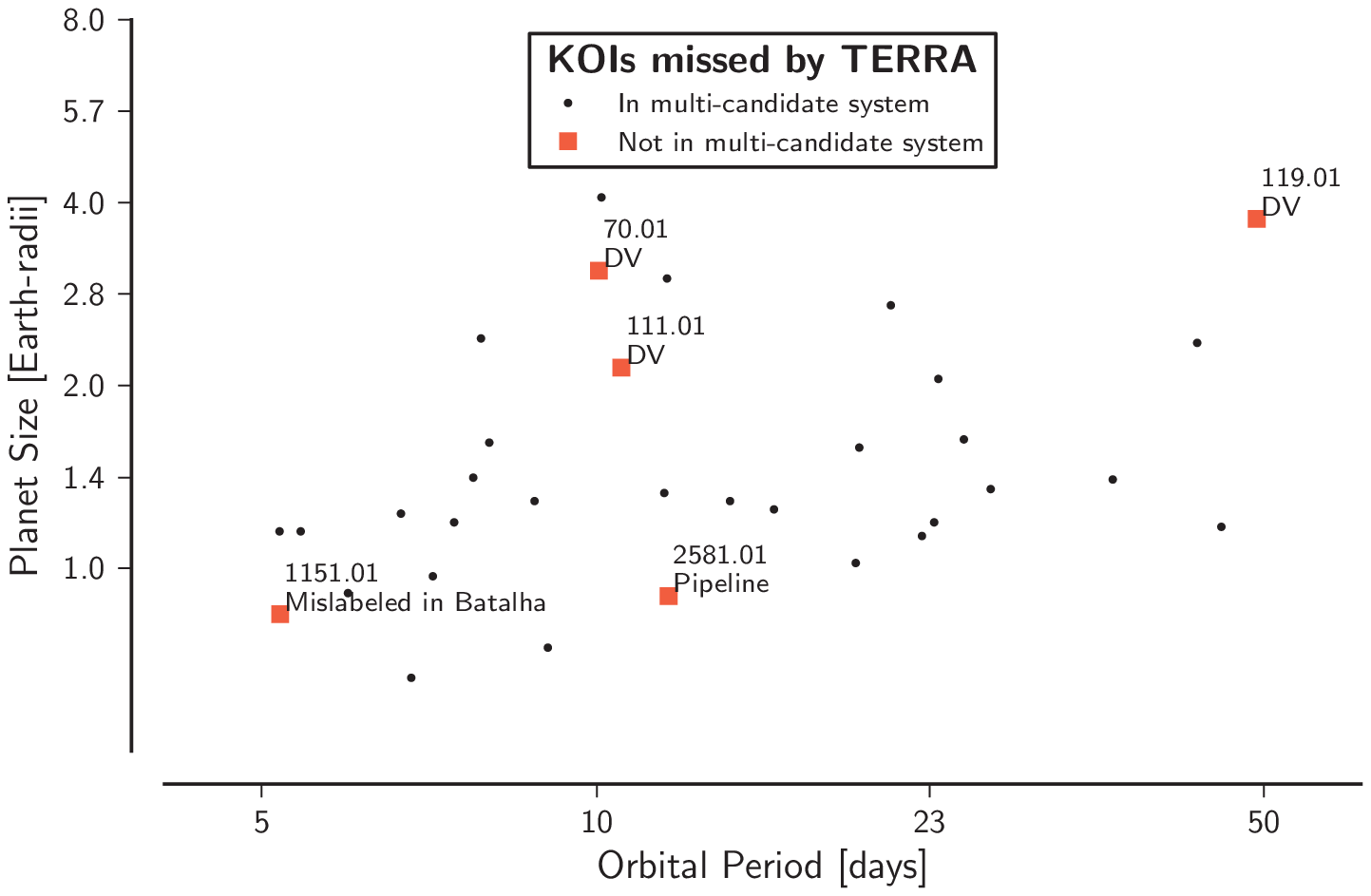}
\caption{\Per and \Rp for the 33 candidates present in \cite{Batalha12} but not found by \TERRA. The small symbols show the candidates in mulit-planet systems. \TERRA is blind to such candidates. The 5 larger symbols show the other failure modes of \TERRA: 2581.01 failed due to a pipeline bug; 70.01, 111.01, and 119.01 did not pass DV; and \TERRA missed KOI-1151.01 because it is listed in \cite{Batalha12} with the incorrect period. Most of the missed planets have \Rp < 1.4 \Re.}
\label{fig:TERRAmissed}
\end{figure*}

We plot the 33 total candidates listed in \cite{Batalha12}, but not found by \TERRA in Figure~\ref{fig:TERRAmissed}. We highlight the 5 missed candidates that cannot be explained by the fact that they are a lower SNR candidate in a multi-candidate system. \TERRA is blind to planets in systems with another planet with higher SNR. Figure~\ref{fig:TERRAmissed} shows that most of these missed planets occur at \Rp < 1.4 \Re.

\section{Occurrence with Planet Multiplicity Included}
\label{sec:OccurCOMB}
While the \TERRA planet occurrence measurement benefits from well-characterized completeness, it does not include the contribution of multis to overall planet occurrence. As discussed in Sections~\ref{sec:TERRAgrid} and \ref{sec:catonly}, \TERRA only detects the highest SNR candidate for a given star. Here, we present planet occurrence including multis from \cite{Batalha12}. Thus, the occurrence within a bin, $f$, in this section should be interpreted as the {\em average number of planets} per star with \Per = 5--50 days. The additional planets from \cite{Batalha12} raise the occurrence values somewhat over those of the previous section. However, the rise and plateau structure remains the same.

We compute $\fcell$ from the 32 candidates present in \cite{Batalha12}, but not found by \TERRA (mislabeled KOI-1151.01 was not included). For clarity, we refer to this separate occurrence calculation as \fcellBa. Because the completeness of the \Kepler pipeline is unknown, we apply no completeness correction. This assumption of 100\%  completeness is certainly an overestimate, but we believe that the sensitivity of the \Kepler pipeline to multis is nearly complete for \Rp > 1.4 \Re. \TERRA has > 80\% completeness for \Rp > 1.4 \Re because planets in that size range with \Per = 5--50 days around Best12k stars have high SNR. The \Kepler pipeline should also be detecting these high SNR candidates. Also, once a KOI is found, the \Kepler team reprocesses the light curve for additional transits (Jason Rowe, private communication). Due to this additional scrutiny, we believe that the \Kepler completeness for multis is higher than for singles, all else being equal.

\begin{figure*}
\includegraphics[width=\textwidth]{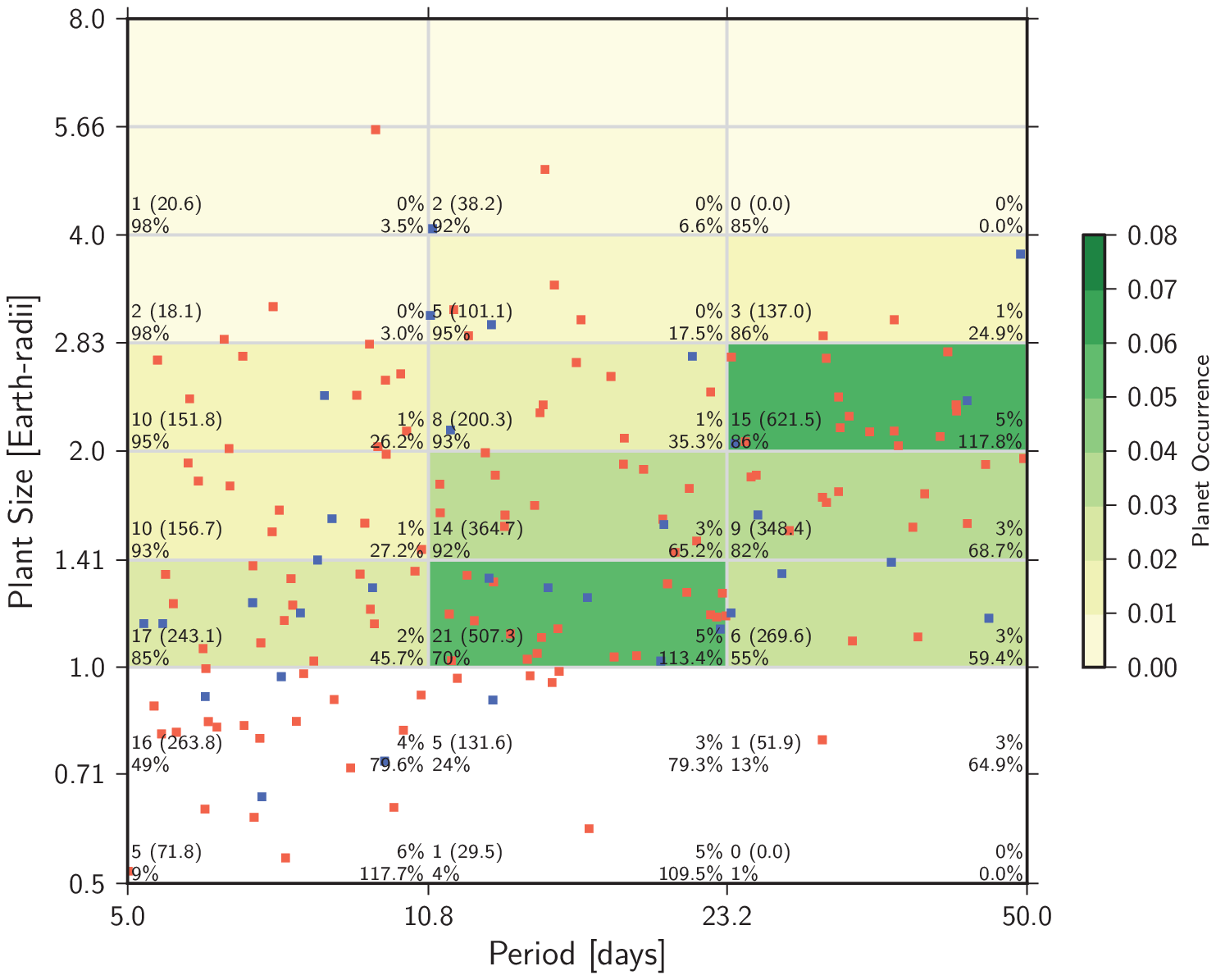}
\caption{As in Figure~\ref{fig:TERRAoccur2D}, red points show 119 \TERRA-detected planets. Blue points represent additional planets from \cite{Batalha12}. Most (28 out of 32) of these new candidates are planets in multi-candidate systems where \TERRA successfully identifies the higher SNR candidate. We apply no completeness correction to these new planets, and we believe this is appropriate for \Rp > 1.4 \Re. We quote the following occurrence information for each cell: Top left--number of planets (number of augmented planets), lower left--completeness, top right--fractional planet occurrence \fcell, bottom right--normalized planet occurrence \flogA. We do not color cells where the completeness is less than 50\% (i.e. the completeness correction is larger than a factor of 2). The planet counts and occurrence values are for the combined \TERRA and \cite{Batalha12} sample. The completeness values are the same as in Figure~\ref{fig:TERRAoccur2D}.}
\label{fig:COMBoccur2D}
\end{figure*}

\begin{figure}
\includegraphics[width=\columnwidth]{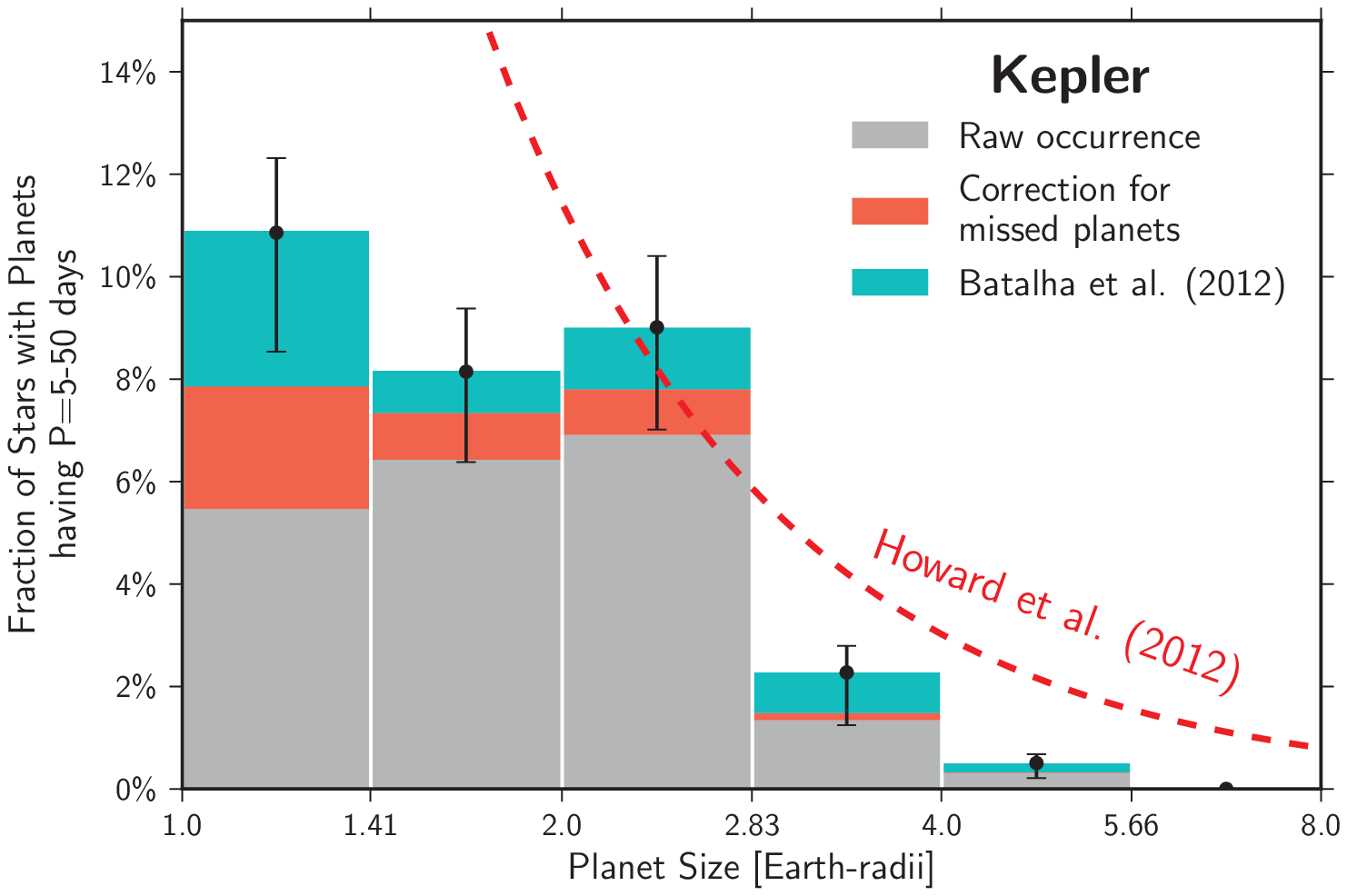}
\caption{Same as Figure~\ref{fig:TERRAoccur1D} with inclusion of planets in multi planet systems. The blue regions represent the additional contribution to planet occurrence from the \cite{Batalha12} planets. The addition of these new planets does not change the overall shape of the distribution. The dashed line is the power law fit to the planet size distribution in H12. The fit agrees qualitatively for \Rp > 2 \Re, but not within errors. We expect the H12 fit to be $\sim25\%$ higher than our occurrence measurements since H12 included planets with \Per~<~50~days (not \Per~=~5--50~days). Additional discrepancies between occurrence in H12 and this work could stem from different characterizations of completeness, reliance on photometric versus spectroscopic measurements of \Rstar, and magnitude-limited, rather than noise-limited, samples.}
\label{fig:COMBoccur1D}
\end{figure}

We then add \fcellBa to $\fcell$ computed in the previous section. We show occurrence computed using \TERRA and \cite{Batalha12} planets as a function of \Per and \Rp in Figure~\ref{fig:COMBoccur2D} and as a function of only \Rp in Figure~\ref{fig:COMBoccur1D}. The 32 additional planets from \cite{Batalha12} do not change the overall shape of the occurrence distribution: rising from 4.0 to 2.8~\Re and consistent with flat from 2.8 down to 1.0~\Re.

H12 fit occurrence for \Rp > 2~\Re with a power law, 
\begin{equation}
\frac{d f }{ d \log \Rp} = k_{R} \Rp^{\alpha}, 
\end{equation}
finding $\alpha = -1.92\pm0.11$ and $k_{R} = 2.9^{+0.5}_{-0.4}$ (Section 3.1 of H12). As a point of comparison, we plot the H12 power law over our combined occurrence distribution in Figure~\ref{fig:COMBoccur1D}. The fit agrees qualitatively for \Rp > 2 \Re, but not within errors. We expect the H12 fit to be $\sim25\%$ higher than our occurrence measurements since H12 included planets with \Per~<~50~days (not \Per~=~5--50~days). Additional discrepancies could stem from  different characterizations of completeness, reliance on photometric versus spectroscopic measurements of \Rstar, and magnitude-limited, rather than noise-limited, samples.  

\section{Discussion}
\label{sec:Discussion}
\subsection{TERRA}

We implement in this work a new pipeline for the detection of transiting planets in \Kepler photometry and apply it to a sample of 12,000 G and K-type dwarfs stars  chosen to be among the most photometrically quiet of the \Kepler target stars.  These low noise stars offer the best chance for the detection of small, Earth-size planets in the \Kepler field and will one day be among the stars from which $\eta_{\oplus}$---the fraction of Sun-like stars bearing Earth-size planets in habitable zone orbits---is estimated. In this work, we focus on the close-in planets having orbital periods of 5--50 days and semi-major axes $\lessapprox$~0.25 AU.   Earth-size planets with these characteristics are statistically at the margins of detectability with the current $\sim$ 3 years of photometry in \Kepler quarters Q1--Q12.  

Our \TERRA pipeline has two key features that enable confident measurement of the occurrence of close-in planets approaching Earth size. First, \TERRA calibrates the \Kepler photometry and searches for transit signals independent of the results from the \Kepler Mission's official pipeline.  In some cases, \TERRA calibration achieves superior noise suppression compared to the Pre-search Data Conditioning (PDC) module of the official \Kepler pipeline \citep{Petigura12}. The transit search algorithm in \TERRA is efficient at detecting low SNR transits in the calibrated light curves.  This algorithm successfully rediscovers 82 of 86 stars bearing planets in \cite{Batalha12}. Recall that the current version of \TERRA only detects the highest SNR transit signal in each system.  Thus, additional planets orbiting known hosts are not reported here.  We report the occurrence of stars having one or more planets, not the mean number of planets per star as in H12 and elsewhere. Our pipeline also detects 37 planets not found in \cite{Batalha12} (19 of which were not in the catalog of the \Kepler team as of August \nth{8}, 2012), albeit with the benefit of 6 quarters of additional photometry for \TERRA to search.  

The second crucial feature of \TERRA is that we have characterized its detection completeness via the injection and recovery of synthetic transits in real \Kepler light curves from the Best12k sample. \textit{This completeness study is crucial to our occurrence calculations because it allows us to statistically correct for incompleteness variations across the \Per--\Rp plane.}  While the \Kepler Project has initiated a completeness study of the official pipeline \citep{Christiansen12}, \TERRA is the only pipeline for \Kepler photometry whose detection completeness has been calibrated by injection and recovery tests.  Prior to \TERRA, occurrence calculations required one to \textit{assume} that the \Kepler planet detections were complete down to some SNR limit, or to estimate completeness based on SNR alone without empirical tests of the performance of the algorithms in the pipeline. For example, H12 made cuts in stellar brightness ($Kp$ < 15) and transit SNR ($>$ 10 in a single quarter of photometry) and restricted their search to planets larger than 2 \Re with orbital periods shorter than 50 days. These conservative cuts on the planet and star catalogs were driven by the unknown completeness of the official \Kepler pipeline at low SNR.  H12 applied two statistical corrections to convert their distribution of detected planets into an occurrence distribution. They corrected for non-transiting planets with a geometric a/\Rstar correction.  They also computed the number of stars amenable to the detection (at SNR~>~10 in a single quarter) of each planet and considered only that number of stars in the occurrence calculation.   H12 had no empirical way to determine the actual detection efficiency of the algorithms in the pipeline. Here, we apply the geometric a/\Rstar correction and correct for pipeline completeness across the \Per--\Rp plane by explicit tests of the \TERRA pipeline efficiency, which naturally incorporates an SNR threshold correction as in H12.  

\subsection{Planet Occurrence}
H12 found that for close-in planets, the planet radius function rises steeply from Jupiter size to 2 \Re.  For smaller planets of $\sim$~1--2 \Re, occurrence was approximately constant in logarithmic \Rp bins, but H12 were skeptical of the result below 2 \Re because of unknown pipeline completeness and the small number statistics near 1 \Re in the \cite{Borucki11} planet catalog. In this work, we strongly confirm the power law rise in occurrence from 4 to 2 \Re using a superior assessment of completeness and nine times more photometry than in H12.  Using \TERRA, we can empirically and confidently compute occurrence down to 1~\Re. Our key result is the plateau of planet occurrence for the size range 1--2.8 \Re for planets having orbital periods 5--50~days around Sun-like stars.  In that size range of 1--2.8~\Re, 23\% of stars have a planet orbiting with periods between 5 and 50 days.  Including the multiple planets within each system, we find 0.28 planets per star within the size range 1--2.8~\Re and with periods between 5--50~days. These results apply, of course, to the \Kepler field, with its still unknown distribution of masses, ages, and metallicities in the Galactic disk.

As shown in Figure~\ref{fig:TERRAoccur2D}, \TERRA detects many sub-Earth size planets (<~1.0~\Re). These sub-Earths appear in regions of low completeness, and, provocatively, may represent just the tip of the iceberg. A rich population of sub-Earths may await discovery given more photometry and continued pipeline improvements. With 8~years of total photometry in an extended \Kepler mission (compared to 3 years here), the computational machinery of \TERRA---including its light curve calibration, transit search, and completeness calibration---will enable a measurement of $\eta_{\oplus}$ for habitable zone orbits.  

\subsection{Interpretation}
We are not the first to note the huge population of close-in planets smaller or less massive than Neptune.  Using Doppler surveys, \cite{Howard10} and \cite{Mayor11} showed that the planet mass function rises steeply with decreasing mass, at least for close-in planets. In \Kepler data, the excess of close-in, small planets was obvious in the initial planet catalogs released by the \Kepler Project \citep{Borucki11a,Borucki11}.  H12 characterized the occurrence distribution of these small planets as a function of their size, orbital period, and host star temperature.  These occurrence measurements, based on official \Kepler planet catalogs, were refined and extended by \cite{Youdin11}, \cite{Traub12}, \cite{Dong12}, \cite{Beauge13}, and others.  Our contribution here shows a clear plateau in occurrence in the 1--2.8 \Re size range and certified by an independent search of Kepler photometry using a pipeline calibrated by injection and recovery tests.  The onset of the plateau at $\sim$ 2.8 \Re suggests that there is a preferred size scale for the formation of close-in planets.

H12 and \cite{Youdin11} noted falling planet occurrence for periods shorter than $\sim$ 7 days. We also observe declining planet occurrence for short orbital periods, but find that the transition occurs closer to $\sim$ 10 days. We consider our period distribution to be in qualitative agreement with those of H12 and \cite{Youdin11}. Planet formation and/or migration seems to discourage very close-in planets (\Per $\lessapprox$ 10 days).

Close-in, small planets are now the most abundant planets detected by current transit and Doppler searches, yet they are absent from the solar system. The solar system is devoid of planets between 1 and 3.88~\Re (Earth and Neptune) and planets with periods less than Mercury's (\Per < 88.0 days). The formation mechanisms and possible subsequent migration of such planets are hotly debated.  The population synthesis models of \cite{Ida10} and \cite{Mordasini12} suggest that they form near or beyond the ice line and then migrate quiescently in the protoplanetary disk.  These models follow the growth and migration of planets over a wide range of parameters (from Jupiter mass down to Earth mass orbiting at distances out to $\sim10$ AU) and they predict ``deserts'' of planet occurrence that are not detected.

More recently, \cite{Hansen12} and \cite{Chiang12} have argued for the {\em in situ} formation of close-in planets of Neptune size and smaller.  In these models, close-in rocky planets of a few Earth masses form from protoplanetary disks more massive than the minimum mass solar nebula.  Multiple planets per disk form commonly in these models and accretion is fast ($\sim 10^5$ years) and efficient due to the short dynamical timescales of close-in orbits.  The rocky cores form before the protoplanetary disk has dissipated, accreting nebular gas that adds typically $\sim$ 3\% to the mass of the planet \citep{Chiang12}.  But the small amounts of gas can significantly swell the radii of these otherwise rocky planets.  For example, \cite{Adams08} found that adding a H/He gas envelope equivalent to 0.2--20\% of the mass of a solid 5 $M_E$  planet increases the radius 8--110\% above the gas-free value.  

We find the {\em in situ} model plausible because it naturally explains the large number of close, sub-Neptune-size planets, the high rate of planet multiplicity and nearly co-planar and circular orbits \citep{Lissauer11, Fang12}, and does not require tuning of planet migration models. Our result of a plateau in the planet size distribution for 1--2.8~\Re with a sharp falloff in occurrence for larger planets along with decreasing occurrence for \Per $\lessapprox$ 10~days are two significant observed properties of planets around Sun-like stars that must be reproduced by models that form planets {\em in situ} or otherwise and by associated population synthesis models. 

The {\em in situ} model seems supported by the sheer large occurrence of sub-Neptune-size planets within 0.25~AU. It seems unlikely that all such planets form beyond the snow line at $\sim$ 2~AU, which would require inward migration to within 0.25~AU, but not all the way into the star. Such models of formation beyond the snow line seem to require fine tuning of migration and parking mechanisms, as well as the tuning of available water or gas beyond 2~AU, while avoiding runaway gas accretion toward Jupiter masses. Still, {\em in situ} formation seems to require higher densities than those normally assumed in a minimum mass solar nebula \citep{Chiang12} in order to form the sub-Neptune planets before removal of the gas. If this {\em in situ} model is correct, we expect these sub-Neptune-size planets to be composed of rock plus H and He, rather than rock plus water \citep{Chiang12}. Thus, a test of the {\em in situ} mode of formation involves spectroscopic measurements of the chemical composition of the close-in sub-Neptunes.

\acknowledgements 
The authors are indebted to Jon Jenkins, Stephen Bryson, Howard Isaacson, Rea Kolbl, Jason Rowe, and Eugene Chiang for productive and enlightening conversations that improved this work. We recognize the independent and complementary work of \cite{Fressin13}, who arrive at similar estimates of planet occurrence. We acknowledge salary support for Petigura by the National Science Foundation through the Graduate Research Fellowship Program. This work made use of NASA's Astrophysics Data System Bibliographic Services as well as the {\tt NumPy} \citep{Oliphant07},  {\tt SciPy} \citep{scipy}, {\tt h5py} \citep{Collette08}, {\tt IPython} \citep{ipython}, and {\tt Matplotlib} \citep{matplotlib} Python modules. Finally, we extend special thanks to those of Hawai'ian ancestry on whose sacred mountain of Mauna Kea we are privileged to be guests.  Without their generous hospitality, the Keck observations presented herein would not have been possible.

\bibliographystyle{apj}
\bibliography{eta-earth,eta-earth_hand,eta-earth_papers,new_bibs,python}

\appendix

\LongTables
\begin{deluxetable*}{l *{14}{r} ccc }
\tablecolumns{16} 
\tabletypesize{\scriptsize}
\tablecaption{Planet candidates identified with \TERRA}
\tablewidth{0pt}
\tablehead{
\colhead{} & 
\multicolumn{8}{c}{Light Curve Fit}   &\colhead{}&
\multicolumn{3}{c}{Stellar Parameters} &
\colhead{}\\ 
\cline{2-9}\cline{11-13}\\
	\colhead{kic}		& 
	\colhead{\Per}			&  
    \colhead{$\ep\tablenotemark{a}$}			&  
	\colhead{ $\rratfrac$ }          &
	\colhead{ $\sigma(\rratfrac)$ }	&   
	\colhead{ $\tau$}		&   
	\colhead{ $\sigma(\tau)$}		&   
	\colhead{ $b$ }			 &   
	\colhead{ $\sigma(b)$ }	 &
	\colhead{}           &
	\colhead{\teff} 		&
	\colhead{\logg}		&
	\colhead{\Rstar}		&
	\colhead{\Rp} 		& 
    \colhead{$\sigma(\Rp)$} 		& 
	\colhead{source\tablenotemark{b}}     &
	\colhead{FP}&
	\colhead{B12}     \\
	\colhead{}		& 
	\colhead{(d)}			&  
    \colhead{(d)}			&  
	\colhead{ (\%) }          &
	\colhead{}			&   
	\colhead{ (hrs)}		&   
	\colhead{}		&   
	\colhead{}	    &  
	\colhead{}       &
	\colhead{}       &
	\colhead{(K)} 		&
	\colhead{(cgs)}		&
	\colhead{(\Rsun)}		&
	\colhead{(\Re)}      &
	\colhead{} 		& 
    \colhead{} 		& 
	\colhead{} &
	\colhead{}        \\
}
\startdata
\input{tab1.tex}
\enddata
\tablecomments{Orbital period, \Per; time of transit center, \ep; planet-to-star radius ratio, \rrat; the time for the planet to travel \Rstar during transit, $\tau$; and transit impact parameter, $b$ are all determined from the \cite{Mandel02} light curve fit. By default, stellar parameters \Rstar, \teff, and \logg come from \SpecMatch. If \SpecMatch parameters do not exist, parameters are taken from the corrected KIC values, described in Section~\ref{sec:TERRAplanetYield}. The FP column lists whether a candidate was designated a false positive by the \Kepler team (`Y'--yes, `N'--no, ` '--no designation). The B12 column lists whether a candidate was present in \cite{Batalha12}.}
\tablenotetext{a}{Time of transit center (BJD-2454900).}
\tablenotetext{b}{Source of stellar parameters: `S'--{\tt SpecMatch}-derived parameters using Keck HIRES spectra, `P1'--photometrically-derived parameters from \cite{Batalha12}, `P2'--photometrically-derived parameters computed by the authors. See Section~\ref{sec:TERRAplanetYield} for more details. }
\label{tab:planetsTERRA}
\end{deluxetable*}

\begin{deluxetable}{l *{14}{r} c }
\tablecolumns{14} 
\tabletypesize{\scriptsize}
\setlength{\tabcolsep}{0.05in}
\tablecaption{New candidates identified with \TERRA}
\tablewidth{0pt}
\tablehead{
\colhead{} & 
\multicolumn{8}{c}{Light Curve Fit}   &\colhead{}&
\multicolumn{3}{c}{Stellar Parameters} &
\colhead{}\\ 
\cline{2-9}\cline{11-13}\\
	\colhead{kic}		& 
	\colhead{\Per}			&  
    \colhead{$\ep\tablenotemark{a}$}			&  
	\colhead{ $\rratfrac$ }          &
	\colhead{ $\sigma(\rratfrac)$ }	&   
	\colhead{ $\tau$}		&   
	\colhead{ $\sigma(\tau)$}		&   
	\colhead{ $b$ }			 &   
	\colhead{ $\sigma(b)$ }	 &
	\colhead{}           &
	\colhead{\teff} 		&
	\colhead{\logg}		&
	\colhead{\Rstar}		&
	\colhead{\Rp} 		& 
    \colhead{$\sigma(\Rp)$} 		& 
	\colhead{source\tablenotemark{b}}  \\
	\colhead{}		& 
	\colhead{(d)}			&  
    \colhead{(d)}			&  
	\colhead{ (\%) }          &
	\colhead{}			&   
	\colhead{ (hrs)}		&   
	\colhead{}		&   
	\colhead{}	    &  
	\colhead{}       &
	\colhead{}       &
	\colhead{(K)} 		&
	\colhead{(cgs)}		&
	\colhead{(\Rsun)}		&
	\colhead{(\Re)}      &
	\colhead{} 		& 
    \colhead{} 		\\
}
\startdata
\input{tab3.tex}
\enddata
\tablecomments{The 37 \TERRA candidates not in \cite{Batalha12} and not listed as false positives by the \Kepler team. The column definitions are the same as in Table~\ref{tab:planetsTERRA}}
\label{tab:planetsTERRAnew}
\end{deluxetable}

\begin{deluxetable}{l r r r r r r r}
\tablecolumns{8} 
\tabletypesize{\scriptsize}
\setlength{\tabcolsep}{0.05in}
\tablecaption{Union of \cite{Batalha12} and \TERRA planet candidate catalogs}
\tablewidth{0pt}
\tablehead{
	\colhead{} &
	\multicolumn{3}{c}{Batalha}   &
	\colhead{}&
	\multicolumn{2}{c}{\TERRA}
	\\
	\cline{2-4}\cline{6-7}\\
	\colhead{kic}          & 
	\colhead{KOI}&  
	\colhead{\Per}&
	\colhead{\Rp}&   
	\colhead{}   & 
	\colhead{\Per}  &  
	\colhead{\Rp}&
}
\startdata
\input{tab2.tex}
\enddata
\label{tab:planetsJOIN}
\tablecomments{All \cite{Batalha12} candidates with \Per~=~5--50~days belonging to stars in the Best12k sample are included. Candidates are considered equal if they belong to the same star and the periods in each catalog agree to better than 0.01 days. Eighty-two candidates appear in both catalogs, 33 appear in \cite{Batalha12} only, and 47 appear in this work only (although 10 were listed as false positives by the \Kepler team). Differences in \Rp between the two catalogs stem from different values of \Rstar. Most \TERRA planet candidates have \SpecMatch-derived stellar parameters which are more accurate than \cite{Batalha12} parameters, which were derived from KIC broadband photometry.}
\end{deluxetable}

\begin{figure}[htbp]
\begin{center}
\includegraphics[width=1\textwidth]{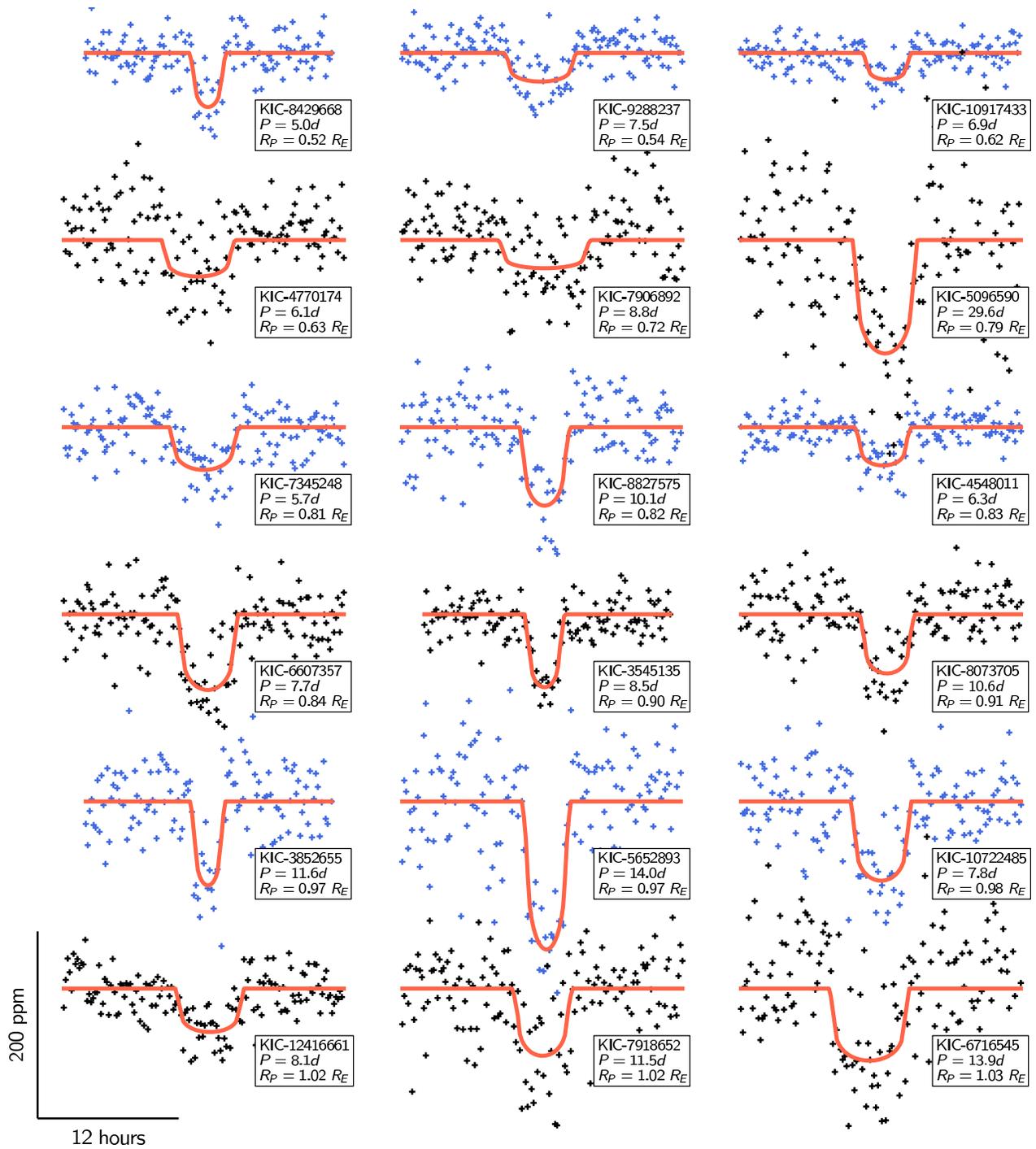}
\caption{Phase-folded photometry for 18 of the 37 \TERRA planet candidates, not in \cite{Batalha12}, ordered according to size. For clarity, we show median photometric measurements in 10~min bins. The red lines are the best-fitting \cite{Mandel02} model.}
\label{fig:binned1}
\end{center}
\end{figure}

\begin{figure}[htbp]
\begin{center}
\includegraphics[width=1\textwidth]{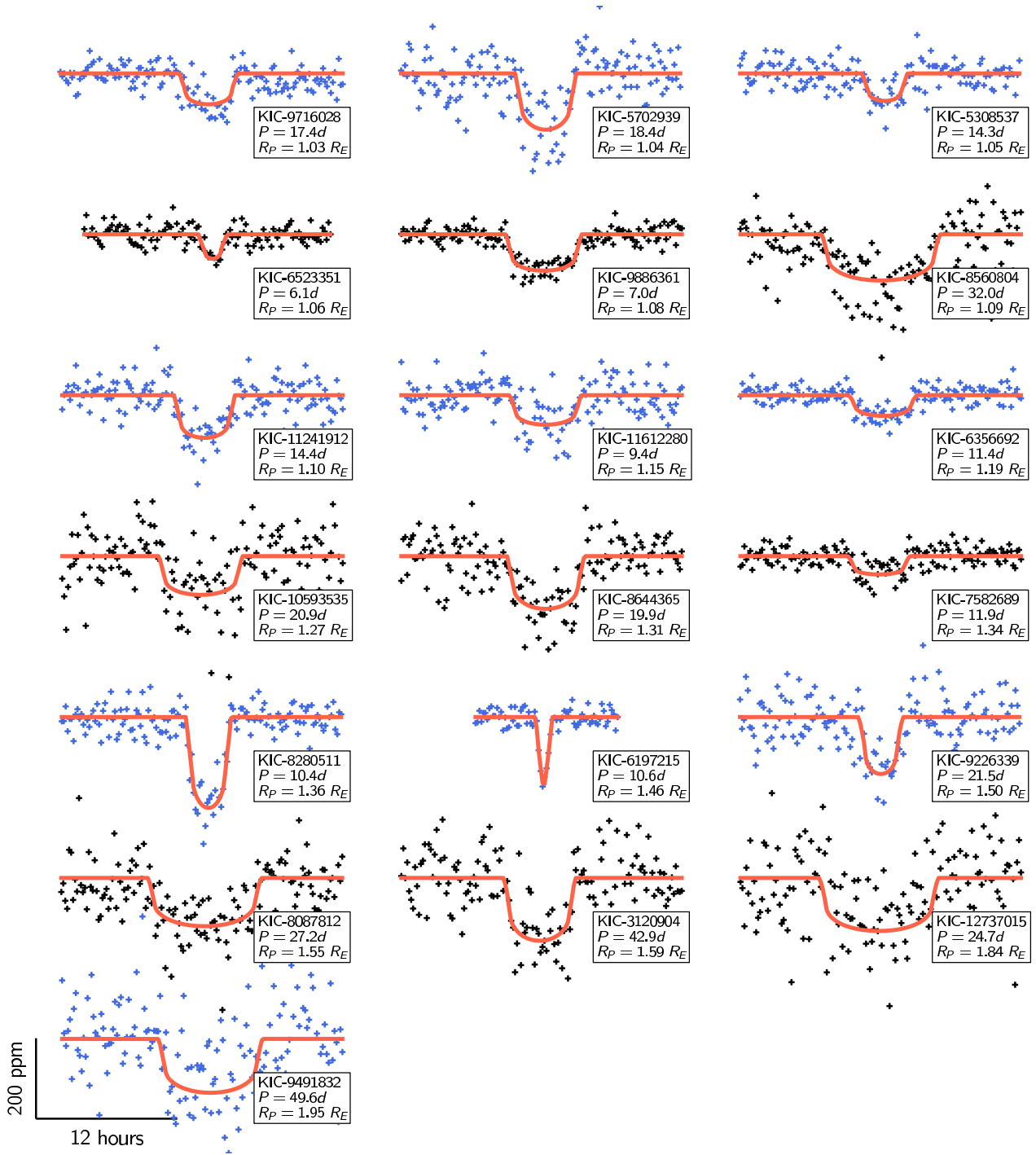}
\caption{Same as Figure~\ref{fig:binned1}, but showing the remaining 19 of the 37 \TERRA planet candidates, not in \cite{Batalha12}.}
\label{fig:binned2}
\end{center}
\end{figure}

\end{document}

%% file: tab1.tex
  2142522 &  13.323 &     67.043 &  0.98 &  0.34 &  2.32 &  0.51 &  < 0.85 &       &        &  6046 &  4.40 &   1.04 &  1.11 &  0.39 &     P1 &     Y &   N \\
  2307415 &  13.122 &     66.396 &  1.26 &  0.13 &  1.90 &  0.12 &  < 0.52 &       &        &  6133 &  4.38 &   1.14 &  1.57 &  0.16 &      S &     N &   Y \\
  2441495 &  12.493 &     71.457 &  2.38 &  0.24 &  1.13 &  0.03 &  < 0.47 &       &        &  5192 &  4.56 &   0.76 &  1.99 &  0.20 &      S &     N &   Y \\
  2444412 &  14.911 &     74.337 &  3.38 &  0.34 &  4.24 &  0.31 &    0.93 &  0.01 &        &  5551 &  4.47 &   0.92 &  3.41 &  0.34 &      S &     N &   Y \\
  2571238 &   9.287 &     68.984 &  2.90 &  0.29 &  3.72 &  0.38 &    0.90 &  0.03 &        &  5544 &  4.50 &   0.89 &  2.82 &  0.28 &      S &     N &   Y \\
  2853446 &   7.373 &     70.613 &  1.39 &  0.14 &  0.74 &  0.06 &  < 0.70 &       &        &  5969 &  4.37 &   1.10 &  1.65 &  0.17 &      S &     N &   Y \\
  3098810 &  40.811 &     75.673 &  1.98 &  0.20 &  1.45 &  0.12 &  < 0.70 &       &        &  6071 &  4.31 &   1.27 &  2.75 &  0.28 &      S &     N &   Y \\
  3120904 &  42.915 &     72.866 &  1.15 &  0.12 &  2.96 &  0.36 &  < 0.66 &       &        &  6151 &  4.31 &   1.26 &  1.59 &  0.16 &      S &       &   N \\
  3342794 &  14.172 &     75.591 &  1.40 &  0.14 &  1.31 &  0.13 &  < 0.55 &       &        &  5900 &  4.35 &   1.10 &  1.68 &  0.17 &      S &     N &   Y \\
  3442055 &  29.619 &     66.681 &  1.57 &  0.16 &  2.48 &  0.15 &  < 0.57 &       &        &  5624 &  4.41 &   1.01 &  1.72 &  0.17 &      S &     N &   Y \\
  3531558 &  24.994 &     71.674 &  1.49 &  0.15 &  2.93 &  0.12 &  < 0.53 &       &        &  5808 &  4.35 &   1.14 &  1.85 &  0.19 &      S &     N &   Y \\
  3545135 &   8.483 &     65.973 &  0.81 &  0.08 &  1.44 &  0.10 &  < 0.57 &       &        &  5794 &  4.40 &   1.02 &  0.90 &  0.09 &      S &     N &   N \\
  3835670 &  14.558 &     78.084 &  2.86 &  0.29 &  3.89 &  0.22 &    0.31 &  0.19 &        &  5722 &  4.14 &   1.58 &  4.93 &  0.49 &      S &     N &   Y \\
  3839488 &  11.131 &     67.370 &  1.36 &  0.14 &  1.97 &  0.11 &  < 0.50 &       &        &  5991 &  4.36 &   1.11 &  1.64 &  0.17 &      S &     N &   Y \\
  3852655 &  11.629 &     65.817 &  0.85 &  0.30 &  1.49 &  0.23 &  < 0.79 &       &        &  5822 &  4.36 &   1.05 &  0.97 &  0.34 &     P1 &     N &   N \\
  3942670 &  33.416 &     70.911 &  1.48 &  0.15 &  3.99 &  0.24 &  < 0.63 &       &        &  6012 &  4.28 &   1.32 &  2.13 &  0.21 &      S &     N &   Y \\
  4043190 &   6.401 &     69.883 &  1.07 &  0.11 &  1.35 &  0.08 &  < 0.65 &       &        &  5302 &  3.83 &   2.45 &  2.86 &  0.29 &      S &     N &   Y \\
  4049901 &  16.291 &     65.115 &  0.64 &  0.23 &  1.72 &  0.16 &  < 0.68 &       &        &  5250 &  4.48 &   0.85 &  0.60 &  0.21 &     P1 &     N &   Y \\
  4548011 &   6.284 &     66.198 &  0.60 &  0.06 &  2.23 &  0.64 &  < 0.49 &       &        &  5991 &  4.30 &   1.26 &  0.83 &  0.09 &      S &       &   N \\
  4644604 &  14.486 &     64.550 &  2.03 &  0.21 &  1.49 &  0.15 &  < 0.59 &       &        &  5739 &  4.34 &   1.04 &  2.32 &  0.23 &      S &     N &   Y \\
  4770174 &   6.096 &     67.600 &  0.57 &  0.20 &  3.08 &  0.44 &  < 0.75 &       &        &  6013 &  4.44 &   1.01 &  0.63 &  0.22 &     P1 &     N &   N \\
  4827723 &   7.239 &     68.024 &  1.63 &  0.17 &  2.20 &  0.37 &    0.70 &  0.12 &        &  5392 &  4.52 &   0.87 &  1.54 &  0.16 &      S &     N &   Y \\
  4914423 &  15.965 &     75.182 &  2.17 &  0.22 &  3.14 &  0.28 &    0.71 &  0.06 &        &  5904 &  4.27 &   1.29 &  3.05 &  0.31 &      S &     N &   Y \\
  4914566 &  22.241 &     77.736 &  0.83 &  0.29 &  3.76 &  0.36 &  < 0.73 &       &        &  5974 &  4.22 &   1.31 &  1.18 &  0.41 &     P1 &     Y &   N \\
  5009743 &  41.699 &    102.563 &  1.94 &  0.20 &  2.87 &  0.30 &  < 0.70 &       &        &  5937 &  4.35 &   1.09 &  2.32 &  0.23 &      S &     N &   Y \\
  5042210 &  12.147 &     64.535 &  0.81 &  0.08 &  3.22 &  0.21 &  < 0.56 &       &        &  6007 &  4.27 &   1.31 &  1.16 &  0.12 &      S &     N &   Y \\
  5094751 &   6.482 &     68.943 &  1.69 &  0.17 &  2.51 &  0.31 &    0.69 &  0.13 &        &  5929 &  4.37 &   1.10 &  2.02 &  0.20 &      S &     N &   Y \\
  5096590 &  29.610 &     70.332 &  1.00 &  0.35 &  2.60 &  0.20 &  < 0.67 &       &        &  5623 &  4.63 &   0.73 &  0.79 &  0.28 &     P1 &       &   N \\
  5121511 &  30.996 &     93.790 &  2.35 &  0.24 &  1.40 &  0.06 &  < 0.47 &       &        &  5217 &  4.53 &   0.84 &  2.15 &  0.22 &      S &     N &   Y \\
  5308537 &  14.265 &     76.217 &  0.76 &  0.08 &  1.92 &  0.53 &  < 0.87 &       &        &  5831 &  4.29 &   1.26 &  1.05 &  0.11 &      S &       &   N \\
  5561278 &  20.310 &     79.826 &  1.16 &  0.12 &  2.73 &  0.15 &  < 0.47 &       &        &  6161 &  4.38 &   1.14 &  1.45 &  0.15 &      S &     N &   Y \\
  5613330 &  23.449 &     69.031 &  1.68 &  0.17 &  4.44 &  0.28 &    0.25 &  0.22 &        &  6080 &  4.20 &   1.48 &  2.70 &  0.27 &      S &     N &   Y \\
  5652893 &  14.010 &     65.959 &  1.15 &  0.13 &  1.98 &  0.49 &  < 0.85 &       &        &  5150 &  4.55 &   0.78 &  0.97 &  0.11 &      S &     N &   N \\
  5702939 &  18.398 &     77.934 &  1.09 &  0.38 &  2.45 &  0.27 &  < 0.54 &       &        &  5634 &  4.47 &   0.87 &  1.04 &  0.36 &     P1 &     N &   N \\
  5735762 &   9.674 &     68.009 &  2.73 &  0.27 &  1.61 &  0.06 &  < 0.47 &       &        &  5195 &  4.53 &   0.84 &  2.51 &  0.25 &      S &     N &   Y \\
  5866724 &   5.860 &     65.040 &  1.65 &  0.17 &  2.01 &  0.05 &    0.22 &  0.14 &        &  6109 &  4.29 &   1.31 &  2.37 &  0.24 &      S &     N &   Y \\
  5959719 &   6.738 &     66.327 &  0.99 &  0.10 &  1.30 &  0.21 &  < 0.76 &       &        &  5166 &  4.56 &   0.77 &  0.83 &  0.09 &      S &     N &   Y \\
  6071903 &  24.308 &     87.054 &  2.28 &  0.23 &  1.29 &  0.05 &  < 0.27 &       &        &  5296 &  4.55 &   0.83 &  2.06 &  0.21 &      S &     N &   Y \\
  6197215 &  10.613 &     68.691 &  1.23 &  0.13 &  0.48 &  0.06 &  < 0.80 &       &        &  5933 &  4.39 &   1.09 &  1.46 &  0.15 &      S &     N &   N \\
  6289257 &  19.675 &     69.903 &  1.33 &  0.13 &  1.96 &  0.21 &    0.35 &  0.27 &        &  6023 &  4.36 &   1.11 &  1.61 &  0.16 &      S &     N &   Y \\
  6291837 &  35.596 &     84.942 &  2.45 &  0.25 &  3.18 &  0.42 &  < 0.71 &       &        &  6165 &  4.38 &   1.14 &  3.05 &  0.31 &      S &     N &   Y \\
  6356692 &  11.392 &     74.555 &  0.66 &  0.23 &  2.74 &  0.31 &  < 0.73 &       &        &  5420 &  4.03 &   1.64 &  1.19 &  0.42 &     P1 &     N &   N \\
  6365156 &  10.214 &     73.050 &  1.57 &  0.16 &  2.83 &  0.09 &  < 0.42 &       &        &  5852 &  4.28 &   1.25 &  2.13 &  0.21 &      S &     N &   Y \\
  6442340 &  13.137 &     76.973 &  1.38 &  0.14 &  2.41 &  0.13 &  < 0.55 &       &        &  5764 &  4.38 &   1.08 &  1.63 &  0.16 &      S &     N &   Y \\
  6521045 &  12.816 &     68.772 &  1.37 &  0.14 &  3.42 &  0.34 &    0.40 &  0.19 &        &  5874 &  4.29 &   1.24 &  1.85 &  0.19 &      S &     N &   Y \\
  6523351 &   6.067 &     69.128 &  0.72 &  0.25 &  1.02 &  0.20 &  < 0.89 &       &        &  5489 &  4.15 &   1.35 &  1.06 &  0.37 &     P1 &       &   N \\
  6605493 &   9.310 &     69.379 &  0.99 &  0.10 &  1.80 &  0.16 &  < 0.62 &       &        &  5805 &  4.36 &   1.12 &  1.20 &  0.12 &      S &     N &   Y \\
  6607357 &   7.700 &     67.390 &  0.85 &  0.30 &  2.72 &  0.45 &    0.55 &  0.22 &        &  5592 &  4.51 &   0.91 &  0.84 &  0.29 &     P1 &     N &   N \\
  6707835 &  22.248 &     84.881 &  2.25 &  0.23 &  1.97 &  0.07 &  < 0.43 &       &        &  5619 &  4.44 &   0.99 &  2.42 &  0.24 &      S &     N &   Y \\
  6716545 &  13.910 &     75.855 &  0.84 &  0.30 &  3.29 &  0.68 &  < 0.85 &       &        &  6044 &  4.30 &   1.12 &  1.03 &  0.36 &     P1 &     N &   N \\
  6803202 &  21.061 &     76.591 &  1.58 &  0.16 &  2.64 &  0.08 &  < 0.36 &       &        &  5719 &  4.40 &   1.03 &  1.77 &  0.18 &      S &     N &   Y \\
  6851425 &  11.120 &     72.744 &  2.29 &  0.23 &  1.71 &  0.12 &  < 0.62 &       &        &  5071 &  4.59 &   0.72 &  1.80 &  0.18 &      S &     N &   Y \\
  6922710 &  23.127 &     78.663 &  1.01 &  0.11 &  2.57 &  0.47 &  < 0.78 &       &        &  5929 &  4.40 &   1.07 &  1.18 &  0.12 &      S &     N &   Y \\
  7021534 &   9.066 &     68.147 &  1.42 &  0.50 &  0.81 &  0.05 &  < 0.55 &       &        &  5848 &  4.55 &   0.87 &  1.35 &  0.47 &     P1 &     Y &   N \\
  7033671 &   9.490 &     66.961 &  1.48 &  0.15 &  1.81 &  0.09 &  < 0.46 &       &        &  5679 &  4.29 &   1.26 &  2.03 &  0.20 &      S &     N &   Y \\
  7211221 &   5.621 &     69.903 &  1.21 &  0.12 &  1.23 &  0.11 &  < 0.53 &       &        &  5634 &  4.44 &   0.93 &  1.23 &  0.12 &      S &     N &   Y \\
  7219825 &  17.233 &     68.084 &  2.06 &  0.21 &  2.27 &  0.13 &  < 0.22 &       &        &  6089 &  4.36 &   1.13 &  2.54 &  0.25 &      S &     N &   Y \\
  7345248 &   5.665 &     69.338 &  0.63 &  0.22 &  2.81 &  0.43 &  < 0.61 &       &        &  5656 &  4.27 &   1.19 &  0.81 &  0.29 &     P2 &       &   N \\
  7419318 &  18.736 &     73.153 &  2.13 &  0.22 &  1.47 &  0.11 &  < 0.61 &       &        &  5187 &  4.54 &   0.81 &  1.89 &  0.19 &      S &     N &   Y \\
  7466863 &  11.971 &     68.173 &  1.83 &  0.64 &  0.98 &  0.03 &  < 0.39 &       &        &  6035 &  4.12 &   1.45 &  2.89 &  1.01 &     P1 &     Y &   N \\
  7582689 &  11.921 &     70.475 &  0.75 &  0.26 &  5.84 &  1.24 &  < 0.95 &       &        &  6022 &  4.04 &   1.64 &  1.34 &  0.47 &     P1 &       &   N \\
  7668663 &   6.498 &     69.012 &  1.44 &  0.15 &  1.85 &  0.34 &    0.79 &  0.09 &        &  5725 &  4.33 &   1.14 &  1.79 &  0.18 &      S &     N &   Y \\
  7700622 &  35.585 &     86.426 &  2.83 &  0.28 &  1.99 &  0.10 &  < 0.46 &       &        &  4787 &  4.62 &   0.69 &  2.13 &  0.21 &      S &     N &   Y \\
  7762723 &   9.887 &     72.539 &  0.73 &  0.26 &  2.21 &  0.39 &  < 0.65 &       &        &  5501 &  4.58 &   0.80 &  0.64 &  0.22 &     P1 &     Y &   N \\
  7810483 &  29.921 &     79.024 &  1.71 &  0.60 &  1.58 &  0.11 &  < 0.58 &       &        &  5893 &  4.52 &   0.91 &  1.70 &  0.59 &     P1 &     Y &   N \\
  7906739 &   7.015 &     69.903 &  0.90 &  0.31 &  2.43 &  0.32 &  < 0.68 &       &        &  5652 &  4.53 &   0.81 &  0.80 &  0.28 &     P1 &     Y &   N \\
  7906892 &   8.849 &     72.797 &  0.58 &  0.07 &  8.03 &  3.69 &    0.89 &  0.12 &        &  6095 &  4.35 &   1.14 &  0.72 &  0.08 &      S &       &   N \\
  7918652 &  11.456 &     69.316 &  0.79 &  0.28 &  2.45 &  0.47 &  < 0.81 &       &        &  5809 &  4.25 &   1.19 &  1.02 &  0.36 &     P1 &     N &   N \\
  8008067 &  15.771 &     70.584 &  2.22 &  0.22 &  3.32 &  0.31 &    0.67 &  0.08 &        &  5594 &  4.37 &   1.10 &  2.66 &  0.27 &      S &     N &   Y \\
  8009496 &  38.476 &     83.567 &  1.88 &  0.66 &  1.72 &  0.14 &  < 0.70 &       &        &  5833 &  4.54 &   0.85 &  1.74 &  0.61 &     P1 &     Y &   N \\
  8073705 &  10.601 &     65.967 &  0.75 &  0.08 &  2.29 &  0.36 &  < 0.80 &       &        &  6086 &  4.36 &   1.12 &  0.91 &  0.10 &      S &       &   N \\
  8077137 &  15.090 &     78.772 &  0.78 &  0.08 &  2.47 &  0.65 &  < 0.85 &       &        &  6179 &  4.37 &   1.16 &  0.99 &  0.10 &      S &     N &   Y \\
  8081187 &  37.323 &     85.828 &  1.67 &  0.59 &  3.58 &  0.39 &  < 0.62 &       &        &  6030 &  4.55 &   0.86 &  1.57 &  0.55 &     P1 &     Y &   N \\
  8087812 &  27.211 &     65.360 &  1.01 &  0.10 &  5.01 &  0.57 &  < 0.67 &       &        &  5985 &  4.17 &   1.41 &  1.55 &  0.16 &      S &       &   N \\
  8242434 &  44.964 &     77.565 &  2.58 &  0.26 &  2.58 &  0.15 &    0.31 &  0.22 &        &  4692 &  4.63 &   0.68 &  1.92 &  0.19 &      S &     N &   Y \\
  8280511 &  10.435 &     67.826 &  1.37 &  0.14 &  1.70 &  0.12 &  < 0.58 &       &        &  5522 &  4.45 &   0.91 &  1.36 &  0.14 &      S &       &   N \\
  8323753 &   6.714 &     67.308 &  1.97 &  0.69 &  1.33 &  0.05 &  < 0.52 &       &        &  5817 &  4.23 &   1.26 &  2.71 &  0.95 &     P1 &     Y &   N \\
  8349582 &  11.523 &     64.959 &  2.14 &  0.22 &  2.41 &  0.21 &    0.61 &  0.09 &        &  5668 &  4.23 &   1.35 &  3.15 &  0.32 &      S &     N &   Y \\
  8429668 &   5.007 &     67.696 &  0.73 &  0.26 &  1.66 &  0.47 &    0.62 &  0.29 &        &  5034 &  4.63 &   0.65 &  0.52 &  0.18 &     P2 &       &   N \\
  8480285 &  29.667 &     92.687 &  2.43 &  0.25 &  4.72 &  0.42 &    0.48 &  0.14 &        &  5960 &  4.36 &   1.09 &  2.89 &  0.29 &      S &     N &   Y \\
  8494617 &  22.923 &     66.640 &  1.06 &  0.11 &  3.72 &  0.36 &  < 0.50 &       &        &  5905 &  4.36 &   1.10 &  1.27 &  0.13 &      S &     N &   Y \\
  8560804 &  31.976 &     66.803 &  0.99 &  0.35 &  4.96 &  0.48 &  < 0.65 &       &        &  5878 &  4.44 &   1.01 &  1.09 &  0.38 &     P1 &     N &   N \\
  8611832 &  22.597 &     73.431 &  1.08 &  0.11 &  3.15 &  0.17 &  < 0.58 &       &        &  5577 &  4.37 &   0.99 &  1.17 &  0.12 &      S &     N &   Y \\
  8628758 &  14.374 &     71.204 &  1.99 &  0.20 &  4.76 &  0.68 &  < 0.91 &       &        &  5773 &  4.40 &   1.04 &  2.26 &  0.23 &      S &     N &   Y \\
  8631504 &  14.820 &     66.409 &  1.20 &  0.12 &  2.02 &  0.25 &  < 0.63 &       &        &  4828 &  4.60 &   0.72 &  0.95 &  0.10 &      S &     N &   Y \\
  8644365 &  19.917 &     72.354 &  1.07 &  0.11 &  3.37 &  0.55 &  < 0.85 &       &        &  6054 &  4.39 &   1.12 &  1.31 &  0.13 &      S &       &   N \\
  8804455 &   7.597 &     64.958 &  1.14 &  0.12 &  2.55 &  0.47 &    0.74 &  0.11 &        &  5715 &  4.38 &   1.07 &  1.33 &  0.14 &      S &     N &   Y \\
  8805348 &  29.907 &     78.383 &  2.36 &  0.24 &  3.20 &  0.39 &    0.66 &  0.12 &        &  5739 &  4.34 &   1.05 &  2.69 &  0.27 &      S &     N &   Y \\
  8822366 &  30.864 &     65.012 &  1.41 &  0.14 &  4.16 &  0.20 &  < 0.62 &       &        &  6089 &  4.35 &   1.14 &  1.76 &  0.18 &      S &     N &   Y \\
  8827575 &  10.129 &     68.997 &  0.84 &  0.30 &  1.94 &  0.18 &  < 0.63 &       &        &  5284 &  4.45 &   0.89 &  0.82 &  0.29 &     P1 &       &   N \\
  8866102 &  17.834 &    114.225 &  1.66 &  0.17 &  2.28 &  0.06 &  < 0.39 &       &        &  6178 &  4.37 &   1.15 &  2.08 &  0.21 &      S &     N &   Y \\
  8962094 &  30.865 &     75.052 &  2.09 &  0.21 &  1.37 &  0.10 &  < 0.61 &       &        &  5739 &  4.34 &   1.04 &  2.38 &  0.24 &      S &     N &   Y \\
  8972058 &   8.991 &     69.746 &  2.01 &  0.20 &  2.10 &  0.11 &  < 0.52 &       &        &  5979 &  4.38 &   1.09 &  2.39 &  0.24 &      S &     N &   Y \\
  9006186 &   5.453 &     67.152 &  0.85 &  0.09 &  1.07 &  0.08 &  < 0.61 &       &        &  5404 &  4.53 &   0.87 &  0.81 &  0.08 &      S &     N &   Y \\
  9086251 &   6.892 &     64.632 &  0.87 &  0.09 &  0.93 &  0.10 &  < 0.68 &       &        &  6044 &  4.22 &   1.45 &  1.38 &  0.14 &      S &     N &   Y \\
  9139084 &   5.836 &     67.853 &  2.07 &  0.21 &  1.06 &  0.03 &  < 0.41 &       &        &  5411 &  4.53 &   0.85 &  1.92 &  0.19 &      S &     N &   Y \\
  9226339 &  21.461 &     65.230 &  1.10 &  0.11 &  1.66 &  0.16 &  < 0.70 &       &        &  5807 &  4.28 &   1.25 &  1.50 &  0.15 &      S &       &   N \\
  9288237 &   7.491 &     68.329 &  0.52 &  0.18 &  3.01 &  0.73 &  < 0.84 &       &        &  5946 &  4.44 &   0.96 &  0.54 &  0.19 &     P1 &       &   N \\
  9491832 &  49.565 &    103.693 &  1.14 &  0.12 &  6.43 &  1.49 &    0.74 &  0.25 &        &  5821 &  4.15 &   1.57 &  1.95 &  0.21 &      S &       &   N \\
  9549648 &   5.992 &     69.883 &  1.46 &  0.15 &  1.80 &  0.45 &    0.89 &  0.06 &        &  6165 &  4.38 &   1.14 &  1.82 &  0.19 &      S &     N &   Y \\
  9704384 &   5.509 &     65.285 &  1.35 &  0.14 &  1.85 &  0.26 &  < 0.75 &       &        &  5448 &  4.50 &   0.91 &  1.35 &  0.14 &      S &     N &   Y \\
  9716028 &  17.373 &     71.258 &  0.82 &  0.08 &  2.46 &  0.27 &  < 0.73 &       &        &  6119 &  4.37 &   1.15 &  1.03 &  0.11 &      S &       &   N \\
  9717943 &   6.110 &     69.903 &  0.72 &  0.08 &  1.80 &  0.87 &    0.79 &  0.20 &        &  5968 &  4.30 &   1.27 &  1.00 &  0.11 &      S &     N &   Y \\
  9886361 &   7.031 &     67.464 &  0.88 &  0.09 &  3.04 &  0.21 &  < 0.42 &       &        &  6090 &  4.39 &   1.13 &  1.08 &  0.11 &      S &     N &   N \\
 10055126 &   9.176 &     71.722 &  1.33 &  0.13 &  2.43 &  0.24 &  < 0.48 &       &        &  5905 &  4.36 &   1.10 &  1.59 &  0.16 &      S &     N &   Y \\
 10130039 &  12.758 &     66.961 &  1.19 &  0.12 &  2.19 &  0.07 &  < 0.38 &       &        &  5828 &  4.42 &   1.01 &  1.31 &  0.13 &      S &     N &   Y \\
 10136549 &   9.693 &     65.809 &  1.14 &  0.12 &  3.44 &  0.44 &    0.57 &  0.20 &        &  5684 &  4.13 &   1.59 &  1.98 &  0.20 &      S &     N &   Y \\
 10212441 &  15.044 &     66.211 &  0.95 &  0.10 &  2.97 &  0.24 &  < 0.57 &       &        &  5939 &  4.35 &   1.09 &  1.13 &  0.11 &      S &     N &   Y \\
 10593535 &  20.925 &     67.900 &  0.93 &  0.10 &  3.63 &  0.59 &  < 0.81 &       &        &  5822 &  4.28 &   1.25 &  1.27 &  0.13 &      S &       &   N \\
 10722485 &   7.849 &     67.907 &  0.87 &  0.09 &  2.61 &  0.51 &  < 0.87 &       &        &  5682 &  4.36 &   1.03 &  0.98 &  0.10 &      S &       &   N \\
 10917433 &   6.912 &     65.190 &  0.51 &  0.05 &  2.03 &  0.57 &  < 0.87 &       &        &  5680 &  4.33 &   1.12 &  0.62 &  0.06 &      S &       &   N \\
 11086270 &  31.720 &     75.821 &  1.90 &  0.19 &  3.35 &  0.40 &    0.70 &  0.14 &        &  5960 &  4.37 &   1.08 &  2.24 &  0.22 &      S &     N &   Y \\
 11121752 &   7.630 &     70.087 &  1.00 &  0.10 &  1.56 &  0.23 &  < 0.75 &       &        &  6045 &  4.36 &   1.12 &  1.22 &  0.12 &      S &     N &   Y \\
 11133306 &  41.746 &    101.657 &  1.90 &  0.19 &  2.32 &  0.15 &    0.27 &  0.22 &        &  5953 &  4.37 &   1.10 &  2.27 &  0.23 &      S &     N &   Y \\
 11241912 &  14.427 &     71.857 &  0.95 &  0.10 &  2.40 &  0.18 &  < 0.59 &       &        &  5931 &  4.40 &   1.07 &  1.10 &  0.11 &      S &       &   N \\
 11250587 &   7.257 &     67.028 &  1.95 &  0.20 &  2.41 &  0.05 &  < 0.40 &       &        &  5853 &  4.18 &   1.49 &  3.18 &  0.32 &      S &     N &   Y \\
 11253711 &  17.791 &     82.259 &  1.86 &  0.65 &  1.29 &  0.09 &  < 0.57 &       &        &  5816 &  4.48 &   0.95 &  1.92 &  0.67 &     P1 &     N &   Y \\
 11295426 &   5.399 &     69.065 &  1.89 &  0.19 &  2.78 &  0.20 &    0.80 &  0.05 &        &  5793 &  4.25 &   1.30 &  2.68 &  0.27 &      S &     N &   Y \\
 11402995 &  10.061 &     71.959 &  2.00 &  0.20 &  2.42 &  0.19 &  < 0.62 &       &        &  5709 &  4.30 &   1.18 &  2.56 &  0.26 &      S &     N &   Y \\
 11554435 &   9.434 &     73.119 &  5.69 &  0.57 &  1.29 &  0.02 &  < 0.43 &       &        &  5536 &  4.52 &   0.90 &  5.60 &  0.56 &      S &     N &   Y \\
 11560897 &  35.968 &     71.667 &  1.46 &  0.15 &  1.47 &  0.11 &  < 0.45 &       &        &  5832 &  4.27 &   1.28 &  2.03 &  0.21 &      S &     N &   Y \\
 11612280 &   9.406 &     70.699 &  0.80 &  0.08 &  3.02 &  0.54 &  < 0.77 &       &        &  5857 &  4.24 &   1.32 &  1.15 &  0.12 &      S &       &   N \\
 11771430 &  40.031 &     81.863 &  1.39 &  0.14 &  2.25 &  0.14 &  < 0.57 &       &        &  5850 &  4.23 &   1.38 &  2.10 &  0.21 &      S &     N &   Y \\
 11774991 &  37.815 &     74.112 &  1.45 &  0.15 &  2.37 &  0.21 &  < 0.65 &       &        &  4710 &  4.62 &   0.70 &  1.10 &  0.11 &      S &     N &   Y \\
 12254909 &   5.350 &     66.872 &  0.84 &  0.29 &  2.26 &  0.14 &  < 0.60 &       &        &  5987 &  4.46 &   0.96 &  0.88 &  0.31 &     P1 &     N &   Y \\
 12301181 &   6.147 &     67.867 &  1.04 &  0.11 &  1.45 &  0.07 &  < 0.40 &       &        &  4997 &  4.60 &   0.74 &  0.84 &  0.09 &      S &     N &   Y \\
 12416661 &   8.053 &     67.968 &  0.63 &  0.22 &  2.93 &  0.37 &  < 0.59 &       &        &  6091 &  4.12 &   1.47 &  1.02 &  0.36 &     P1 &       &   N \\
 12454461 &   7.467 &     69.392 &  0.84 &  0.09 &  1.83 &  0.20 &  < 0.71 &       &        &  6048 &  4.31 &   1.27 &  1.16 &  0.12 &      S &     N &   Y \\
 12737015 &  24.669 &     69.616 &  1.05 &  0.11 &  4.94 &  0.49 &  < 0.69 &       &        &  6045 &  4.15 &   1.60 &  1.84 &  0.19 &      S &       &   N \\

%% file: tab3.tex
  3120904 &  42.915 &     72.866 &  1.15 &  0.12 &  2.96 &  0.36 &  < 0.66 &       &        &  6151 &  4.31 &   1.26 &  1.59 &  0.16 &      S \\
  3545135 &   8.483 &     65.973 &  0.81 &  0.08 &  1.44 &  0.10 &  < 0.57 &       &        &  5794 &  4.40 &   1.02 &  0.90 &  0.09 &      S \\
  3852655 &  11.629 &     65.817 &  0.85 &  0.30 &  1.49 &  0.23 &  < 0.79 &       &        &  5822 &  4.36 &   1.05 &  0.97 &  0.34 &     P1 \\
  4548011 &   6.284 &     66.198 &  0.60 &  0.06 &  2.23 &  0.64 &  < 0.49 &       &        &  5991 &  4.30 &   1.26 &  0.83 &  0.09 &      S \\
  4770174 &   6.096 &     67.600 &  0.57 &  0.20 &  3.08 &  0.44 &  < 0.75 &       &        &  6013 &  4.44 &   1.01 &  0.63 &  0.22 &     P1 \\
  5096590 &  29.610 &     70.332 &  1.00 &  0.35 &  2.60 &  0.20 &  < 0.67 &       &        &  5623 &  4.63 &   0.73 &  0.79 &  0.28 &     P1 \\
  5308537 &  14.265 &     76.217 &  0.76 &  0.08 &  1.92 &  0.53 &  < 0.87 &       &        &  5831 &  4.29 &   1.26 &  1.05 &  0.11 &      S \\
  5652893 &  14.010 &     65.959 &  1.15 &  0.13 &  1.98 &  0.49 &  < 0.85 &       &        &  5150 &  4.55 &   0.78 &  0.97 &  0.11 &      S \\
  5702939 &  18.398 &     77.934 &  1.09 &  0.38 &  2.45 &  0.27 &  < 0.54 &       &        &  5634 &  4.47 &   0.87 &  1.04 &  0.36 &     P1 \\
  6197215 &  10.613 &     68.691 &  1.23 &  0.13 &  0.48 &  0.06 &  < 0.80 &       &        &  5933 &  4.39 &   1.09 &  1.46 &  0.15 &      S \\
  6356692 &  11.392 &     74.555 &  0.66 &  0.23 &  2.74 &  0.31 &  < 0.73 &       &        &  5420 &  4.03 &   1.64 &  1.19 &  0.42 &     P1 \\
  6523351 &   6.067 &     69.128 &  0.72 &  0.25 &  1.02 &  0.20 &  < 0.89 &       &        &  5489 &  4.15 &   1.35 &  1.06 &  0.37 &     P1 \\
  6607357 &   7.700 &     67.390 &  0.85 &  0.30 &  2.72 &  0.45 &    0.55 &  0.22 &        &  5592 &  4.51 &   0.91 &  0.84 &  0.29 &     P1 \\
  6716545 &  13.910 &     75.855 &  0.84 &  0.30 &  3.29 &  0.68 &  < 0.85 &       &        &  6044 &  4.30 &   1.12 &  1.03 &  0.36 &     P1 \\
  7345248 &   5.665 &     69.338 &  0.63 &  0.22 &  2.81 &  0.43 &  < 0.61 &       &        &  5656 &  4.27 &   1.19 &  0.81 &  0.29 &     P2 \\
  7582689 &  11.921 &     70.475 &  0.75 &  0.26 &  5.84 &  1.24 &  < 0.95 &       &        &  6022 &  4.04 &   1.64 &  1.34 &  0.47 &     P1 \\
  7906892 &   8.849 &     72.797 &  0.58 &  0.07 &  8.03 &  3.69 &    0.89 &  0.12 &        &  6095 &  4.35 &   1.14 &  0.72 &  0.08 &      S \\
  7918652 &  11.456 &     69.316 &  0.79 &  0.28 &  2.45 &  0.47 &  < 0.81 &       &        &  5809 &  4.25 &   1.19 &  1.02 &  0.36 &     P1 \\
  8073705 &  10.601 &     65.967 &  0.75 &  0.08 &  2.29 &  0.36 &  < 0.80 &       &        &  6086 &  4.36 &   1.12 &  0.91 &  0.10 &      S \\
  8087812 &  27.211 &     65.360 &  1.01 &  0.10 &  5.01 &  0.57 &  < 0.67 &       &        &  5985 &  4.17 &   1.41 &  1.55 &  0.16 &      S \\
  8280511 &  10.435 &     67.826 &  1.37 &  0.14 &  1.70 &  0.12 &  < 0.58 &       &        &  5522 &  4.45 &   0.91 &  1.36 &  0.14 &      S \\
  8429668 &   5.007 &     67.696 &  0.73 &  0.26 &  1.66 &  0.47 &    0.62 &  0.29 &        &  5034 &  4.63 &   0.65 &  0.52 &  0.18 &     P2 \\
  8560804 &  31.976 &     66.803 &  0.99 &  0.35 &  4.96 &  0.48 &  < 0.65 &       &        &  5878 &  4.44 &   1.01 &  1.09 &  0.38 &     P1 \\
  8644365 &  19.917 &     72.354 &  1.07 &  0.11 &  3.37 &  0.55 &  < 0.85 &       &        &  6054 &  4.39 &   1.12 &  1.31 &  0.13 &      S \\
  8827575 &  10.129 &     68.997 &  0.84 &  0.30 &  1.94 &  0.18 &  < 0.63 &       &        &  5284 &  4.45 &   0.89 &  0.82 &  0.29 &     P1 \\
  9226339 &  21.461 &     65.230 &  1.10 &  0.11 &  1.66 &  0.16 &  < 0.70 &       &        &  5807 &  4.28 &   1.25 &  1.50 &  0.15 &      S \\
  9288237 &   7.491 &     68.329 &  0.52 &  0.18 &  3.01 &  0.73 &  < 0.84 &       &        &  5946 &  4.44 &   0.96 &  0.54 &  0.19 &     P1 \\
  9491832 &  49.565 &    103.693 &  1.14 &  0.12 &  6.43 &  1.49 &    0.74 &  0.25 &        &  5821 &  4.15 &   1.57 &  1.95 &  0.21 &      S \\
  9716028 &  17.373 &     71.258 &  0.82 &  0.08 &  2.46 &  0.27 &  < 0.73 &       &        &  6119 &  4.37 &   1.15 &  1.03 &  0.11 &      S \\
  9886361 &   7.031 &     67.464 &  0.88 &  0.09 &  3.04 &  0.21 &  < 0.42 &       &        &  6090 &  4.39 &   1.13 &  1.08 &  0.11 &      S \\
 10593535 &  20.925 &     67.900 &  0.93 &  0.10 &  3.63 &  0.59 &  < 0.81 &       &        &  5822 &  4.28 &   1.25 &  1.27 &  0.13 &      S \\
 10722485 &   7.849 &     67.907 &  0.87 &  0.09 &  2.61 &  0.51 &  < 0.87 &       &        &  5682 &  4.36 &   1.03 &  0.98 &  0.10 &      S \\
 10917433 &   6.912 &     65.190 &  0.51 &  0.05 &  2.03 &  0.57 &  < 0.87 &       &        &  5680 &  4.33 &   1.12 &  0.62 &  0.06 &      S \\
 11241912 &  14.427 &     71.857 &  0.95 &  0.10 &  2.40 &  0.18 &  < 0.59 &       &        &  5931 &  4.40 &   1.07 &  1.10 &  0.11 &      S \\
 11612280 &   9.406 &     70.699 &  0.80 &  0.08 &  3.02 &  0.54 &  < 0.77 &       &        &  5857 &  4.24 &   1.32 &  1.15 &  0.12 &      S \\
 12416661 &   8.053 &     67.968 &  0.63 &  0.22 &  2.93 &  0.37 &  < 0.59 &       &        &  6091 &  4.12 &   1.47 &  1.02 &  0.36 &     P1 \\
 12737015 &  24.669 &     69.616 &  1.05 &  0.11 &  4.94 &  0.49 &  < 0.69 &       &        &  6045 &  4.15 &   1.60 &  1.84 &  0.19 &      S \\

%% file: tab2.tex
  2142522 &          &        &         &        &  13.32 &   1.11 \\
  2307415 &  2053.01 &  13.12 &    1.65 &        &  13.12 &   1.57 \\
  2441495 &   166.01 &  12.49 &    2.70 &        &  12.49 &   1.99 \\
  2444412 &   103.01 &  14.91 &    2.97 &        &  14.91 &   3.41 \\
  2571238 &    84.01 &   9.29 &    2.53 &        &   9.29 &   2.82 \\
  2853446 &  1118.01 &   7.37 &    2.52 &        &   7.37 &   1.65 \\
  3098810 &  1878.01 &  40.81 &    3.24 &        &  40.81 &   2.75 \\
  3120904 &          &        &         &        &  42.91 &   1.59 \\
  3342794 &  2278.01 &  14.17 &    1.99 &        &  14.17 &   1.68 \\
  3442055 &  1218.01 &  29.62 &    2.22 &        &  29.62 &   1.72 \\
  3531558 &   118.01 &  24.99 &    1.41 &        &  24.99 &   1.85 \\
  3545135 &          &        &         &        &   8.48 &   0.90 \\
  3835670 &   149.01 &  14.56 &    5.50 &        &  14.56 &   4.93 \\
  3839488 &  1216.01 &  11.13 &    1.57 &        &  11.13 &   1.64 \\
  3852655 &          &        &         &        &  11.63 &   0.97 \\
  3942670 &   392.02 &  12.61 &    1.33 &        &        &        \\
  3942670 &   392.01 &  33.42 &    2.27 &        &  33.42 &   2.13 \\
  4043190 &  1220.01 &   6.40 &    1.95 &        &   6.40 &   2.86 \\
  4049901 &  2295.01 &  16.29 &    0.63 &        &  16.29 &   0.60 \\
  4548011 &          &        &         &        &   6.28 &   0.83 \\
  4644604 &   628.01 &  14.49 &    1.87 &        &  14.49 &   2.32 \\
  4770174 &          &        &         &        &   6.10 &   0.63 \\
  4827723 &   632.01 &   7.24 &    1.46 &        &   7.24 &   1.54 \\
  4914423 &   108.01 &  15.97 &    2.94 &        &  15.97 &   3.05 \\
  4914566 &          &        &         &        &  22.24 &   1.18 \\
  5009743 &  1609.01 &  41.70 &    2.34 &        &  41.70 &   2.32 \\
  5042210 &  2462.01 &  12.15 &    1.37 &        &  12.15 &   1.16 \\
  5094751 &   123.01 &   6.48 &    2.64 &        &   6.48 &   2.02 \\
  5094751 &   123.02 &  21.22 &    2.71 &        &        &        \\
  5096590 &          &        &         &        &  29.61 &   0.79 \\
  5121511 &   640.01 &  31.00 &    2.43 &        &  31.00 &   2.15 \\
  5308537 &          &        &         &        &  14.27 &   1.05 \\
  5446285 &   142.01 &  10.92 &    4.08 &        &        &        \\
  5561278 &  1621.01 &  20.31 &    2.48 &        &  20.31 &   1.45 \\
  5613330 &   649.01 &  23.45 &    2.31 &        &  23.45 &   2.70 \\
  5652893 &          &        &         &        &  14.01 &   0.97 \\
  5702939 &          &        &         &        &  18.40 &   1.04 \\
  5735762 &   148.02 &   9.67 &    3.14 &        &   9.67 &   2.51 \\
  5735762 &   148.03 &  42.90 &    2.35 &        &        &        \\
  5866724 &    85.01 &   5.86 &    2.35 &        &   5.86 &   2.37 \\
  5866724 &    85.03 &   8.13 &    1.41 &        &        &        \\
  5959719 &  2498.01 &   6.74 &    0.78 &        &   6.74 &   0.83 \\
  6061773 &  2001.01 &   8.28 &    2.39 &        &        &        \\
  6071903 &   306.01 &  24.31 &    2.28 &        &  24.31 &   2.06 \\
  6197215 &          &        &         &        &  10.61 &   1.46 \\
  6289257 &   307.02 &   5.21 &    1.15 &        &        &        \\
  6289257 &   307.01 &  19.67 &    1.80 &        &  19.67 &   1.61 \\
  6291837 &   308.01 &  35.60 &    3.15 &        &  35.60 &   3.05 \\
  6356692 &          &        &         &        &  11.39 &   1.19 \\
  6365156 &   662.01 &  10.21 &    2.05 &        &  10.21 &   2.13 \\
  6442340 &   664.02 &   7.78 &    1.19 &        &        &        \\
  6442340 &   664.01 &  13.14 &    1.83 &        &  13.14 &   1.63 \\
  6442340 &   664.03 &  23.44 &    1.19 &        &        &        \\
  6521045 &    41.02 &   6.89 &    1.23 &        &        &        \\
  6521045 &    41.01 &  12.82 &    2.08 &        &  12.82 &   1.85 \\
  6521045 &    41.03 &  35.33 &    1.40 &        &        &        \\
  6523351 &          &        &         &        &   6.07 &   1.06 \\
  6605493 &  2559.01 &   9.31 &    0.99 &        &   9.31 &   1.20 \\
  6607357 &          &        &         &        &   7.70 &   0.84 \\
  6678383 &   111.01 &  11.43 &    2.14 &        &        &        \\
  6678383 &   111.02 &  23.67 &    2.05 &        &        &        \\
  6707835 &   666.01 &  22.25 &    2.56 &        &  22.25 &   2.42 \\
  6716545 &          &        &         &        &  13.91 &   1.03 \\
  6803202 &   177.01 &  21.06 &    1.84 &        &  21.06 &   1.77 \\
  6850504 &    70.04 &   6.10 &    0.91 &        &        &        \\
  6850504 &    70.01 &  10.85 &    3.09 &        &        &        \\
  6850504 &    70.05 &  19.58 &    1.02 &        &        &        \\
  6851425 &   163.01 &  11.12 &    2.27 &        &  11.12 &   1.80 \\
  6922710 &  2087.01 &  23.13 &    1.54 &        &  23.13 &   1.18 \\
  7021534 &          &        &         &        &   9.07 &   1.35 \\
  7033671 &   670.01 &   9.49 &    1.92 &        &   9.49 &   2.03 \\
  7211221 &  1379.01 &   5.62 &    1.06 &        &   5.62 &   1.23 \\
  7219825 &   238.01 &  17.23 &    2.40 &        &  17.23 &   2.54 \\
  7219825 &   238.02 &  26.69 &    1.35 &        &        &        \\
  7345248 &          &        &         &        &   5.66 &   0.81 \\
  7419318 &   313.02 &   8.44 &    1.61 &        &        &        \\
  7419318 &   313.01 &  18.74 &    2.20 &        &  18.74 &   1.89 \\
  7466863 &          &        &         &        &  11.97 &   2.89 \\
  7582689 &          &        &         &        &  11.92 &   1.34 \\
  7668663 &  1898.01 &   6.50 &    1.50 &        &   6.50 &   1.79 \\
  7700622 &   315.01 &  35.59 &    2.14 &        &  35.59 &   2.13 \\
  7762723 &          &        &         &        &   9.89 &   0.64 \\
  7810483 &          &        &         &        &  29.92 &   1.70 \\
  7906739 &          &        &         &        &   7.01 &   0.80 \\
  7906892 &          &        &         &        &   8.85 &   0.72 \\
  7918652 &          &        &         &        &  11.46 &   1.02 \\
  8008067 &   316.01 &  15.77 &    2.72 &        &  15.77 &   2.66 \\
  8009496 &          &        &         &        &  38.48 &   1.74 \\
  8073705 &          &        &         &        &  10.60 &   0.91 \\
  8077137 &   274.01 &  15.09 &    1.12 &        &  15.09 &   0.99 \\
  8077137 &   274.02 &  22.80 &    1.13 &        &        &        \\
  8081187 &          &        &         &        &  37.32 &   1.57 \\
  8087812 &          &        &         &        &  27.21 &   1.55 \\
  8242434 &  1726.01 &  44.96 &    5.25 &        &  44.96 &   1.92 \\
  8280511 &  1151.01 &   5.22 &    0.84 &        &        &        \\
  8280511 &  1151.02 &   7.41 &    0.97 &        &        &        \\
  8280511 &          &        &         &        &  10.44 &   1.36 \\
  8323753 &          &        &         &        &   6.71 &   2.71 \\
  8349582 &   122.01 &  11.52 &    2.78 &        &  11.52 &   3.15 \\
  8429668 &          &        &         &        &   5.01 &   0.52 \\
  8480285 &   691.02 &  16.23 &    1.25 &        &        &        \\
  8480285 &   691.01 &  29.67 &    2.92 &        &  29.67 &   2.89 \\
  8494617 &  2389.01 &  22.92 &    1.45 &        &  22.92 &   1.27 \\
  8554498 &     5.02 &   7.05 &    0.66 &        &        &        \\
  8560804 &          &        &         &        &  31.98 &   1.09 \\
  8611832 &  2414.01 &  22.60 &    1.03 &        &  22.60 &   1.17 \\
  8611832 &  2414.02 &  45.35 &    1.17 &        &        &        \\
  8628758 &  1279.02 &   9.65 &    0.74 &        &        &        \\
  8628758 &  1279.01 &  14.37 &    1.31 &        &  14.37 &   2.26 \\
  8631504 &  2503.01 &  14.82 &    2.41 &        &  14.82 &   0.95 \\
  8644365 &          &        &         &        &  19.92 &   1.31 \\
  8804455 &  2159.01 &   7.60 &    1.01 &        &   7.60 &   1.33 \\
  8805348 &   695.01 &  29.91 &    2.51 &        &  29.91 &   2.69 \\
  8822366 &  1282.01 &  30.86 &    3.00 &        &  30.86 &   1.76 \\
  8827575 &          &        &         &        &  10.13 &   0.82 \\
  8866102 &    42.01 &  17.83 &    2.71 &        &  17.83 &   2.08 \\
  8962094 &   700.02 &   9.36 &    1.29 &        &        &        \\
  8962094 &   700.03 &  14.67 &    1.29 &        &        &        \\
  8962094 &   700.01 &  30.86 &    2.28 &        &  30.87 &   2.38 \\
  8972058 &   159.01 &   8.99 &    2.70 &        &   8.99 &   2.39 \\
  9006186 &  2169.01 &   5.45 &    1.02 &        &   5.45 &   0.81 \\
  9086251 &  2367.01 &   6.89 &    1.17 &        &   6.89 &   1.38 \\
  9139084 &   323.01 &   5.84 &    2.17 &        &   5.84 &   1.92 \\
  9226339 &          &        &         &        &  21.46 &   1.50 \\
  9288237 &          &        &         &        &   7.49 &   0.54 \\
  9471974 &   119.01 &  49.18 &    3.76 &        &        &        \\
  9491832 &          &        &         &        &  49.57 &   1.95 \\
  9549648 &  1886.01 &   5.99 &    2.45 &        &   5.99 &   1.82 \\
  9704384 &  1913.01 &   5.51 &    1.40 &        &   5.51 &   1.35 \\
  9716028 &          &        &         &        &  17.37 &   1.03 \\
  9717943 &  2273.01 &   6.11 &    1.02 &        &   6.11 &   1.00 \\
  9886361 &          &        &         &        &   7.03 &   1.08 \\
 10055126 &  1608.01 &   9.18 &    1.81 &        &   9.18 &   1.59 \\
 10055126 &  1608.02 &  19.74 &    1.58 &        &        &        \\
 10130039 &  1909.02 &   5.47 &    1.15 &        &        &        \\
 10130039 &  1909.01 &  12.76 &    1.52 &        &  12.76 &   1.31 \\
 10130039 &  1909.03 &  25.10 &    1.63 &        &        &        \\
 10136549 &  1929.01 &   9.69 &    2.00 &        &   9.69 &   1.98 \\
 10212441 &  2342.01 &  15.04 &    1.22 &        &  15.04 &   1.13 \\
 10593535 &          &        &         &        &  20.92 &   1.27 \\
 10722485 &          &        &         &        &   7.85 &   0.98 \\
 10917433 &          &        &         &        &   6.91 &   0.62 \\
 11086270 &   124.01 &  12.69 &    3.00 &        &        &        \\
 11086270 &   124.02 &  31.72 &    3.58 &        &  31.72 &   2.24 \\
 11121752 &  2333.02 &   7.63 &    1.63 &        &   7.63 &   1.22 \\
 11133306 &   276.01 &  41.75 &    2.49 &        &  41.75 &   2.27 \\
 11241912 &          &        &         &        &  14.43 &   1.10 \\
 11250587 &   107.01 &   7.26 &    3.09 &        &   7.26 &   3.18 \\
 11253711 &  1972.01 &  17.79 &    1.93 &        &  17.79 &   1.92 \\
 11295426 &   246.01 &   5.40 &    2.53 &        &   5.40 &   2.68 \\
 11402995 &   173.01 &  10.06 &    2.48 &        &  10.06 &   2.56 \\
 11554435 &    63.01 &   9.43 &    6.30 &        &   9.43 &   5.60 \\
 11560897 &  2365.01 &  35.97 &    1.59 &        &  35.97 &   2.03 \\
 11612280 &          &        &         &        &   9.41 &   1.15 \\
 11771430 &  2582.01 &  40.03 &    1.98 &        &  40.03 &   2.10 \\
 11774991 &  2173.01 &  37.82 &    1.24 &        &  37.82 &   1.10 \\
 11818872 &  2581.01 &  12.74 &    0.90 &        &        &        \\
 12254909 &  2372.01 &   5.35 &    1.11 &        &   5.35 &   0.88 \\
 12301181 &  2059.01 &   6.15 &    0.59 &        &   6.15 &   0.84 \\
 12416661 &          &        &         &        &   8.05 &   1.02 \\
 12454461 &  2463.01 &   7.47 &    1.07 &        &   7.47 &   1.16 \\
 12737015 &          &        &         &        &  24.67 &   1.84 \\